\documentclass[aps,pra,showpacs,twocolumn]{revtex4}
\usepackage{amsmath}
\usepackage{bm}

\begin{document}

\title{High-precision calculations of In~I and Sn~II atomic properties}

\author{ U. I. Safronova }
\email{usafrono@nd.edu}\altaffiliation{ On  leave  from ISAN,
Troitsk, Russia} \affiliation{Physics Department, University of
Nevada, Reno, Nevada
 89557}

\author{M. S. Safronova}
\email{msafrono@udel.edu} \affiliation {Department of Physics and
Astronomy, 217 Sharp Lab,
 University of Delaware, Newark, Delaware 19716}

\author{M. G. Kozlov}
\affiliation {Petersburg Nuclear Physics
Institute, Gatchina 188300, Russia}

\date{\today}
\begin{abstract}
We use all-order relativistic many-body perturbation theory to
study $5s^2nl$ configurations of In~I and Sn~II. Energies,
E1-amplitudes, and hyperfine constants are calculated using
all-order method, which accounts for single and double excitations
of the Dirac-Fock wave functions. A comprehensive review of
experimental and theoretical studies of In~I and Sn~II properties
is given. Our results are compared with other studies were
available.

\end{abstract}
\maketitle

\section{Introduction}

In this work, we present  a systematic calculation of various In~I
and  Sn~II atomic properties  and   study  the importance of the
high-order correlation corrections to those properties  using
relativistic all-order method. Previously these atoms have been
studied in a number of experimental and theoretical papers. First
theoretical studies were published 30 years ago by \citet{migd76}.
They used relativistic semiempirical method including exchange to
calculate the oscillator strengths in In~I for the $5s^25p_j$ -
$5s^2ns_{1/2}$, $5s^26p_j$ - $5s^2ns_{1/2}$, $5s^25p_j$ -
$5s^2nd_{j}$, and $5s^26s_{1/2}$ - $5s^2np_{j}$ transitions.

Later, the oscillator strengths determined from single-configuration
relativistic Hartree-Fock (RHF) calculations were reported by
\citet{migd79} for the lowest $5s^25p_j$--$5s^26s_{1/2}$ and
$5s^25p_j$--$5s^25d_{j'}$ transitions. A quantum defect theory was
used by \citet{gruzdev78} to calculate oscillator strengths $f$
averaged over $j$ in neutral indium. Configuration interaction gf
values for transitions between the $5s^26s_{1/2}$, $5s^2nd_j$, and
$5s^2np_j$ (with $n$ = 5, 6) states  were reported for the indium
isoelectronic sequence up to Ba VIII in Ref.~\cite{migd94}.   A
self-consistent-field method was used to generate one-electron
orbitals. The method used in ~\cite{migd94} included relativistic
effects albeit in an approximate way, and the configuration
interaction scheme accounts for correlation effects ~\cite{migd94}.
Hartree-Fock calculations including relativistic corrections and
configuration interaction in an intermediate coupling scheme were
carried out in Ref.~\cite{alonso-00} to analyze the spectrum of
Sn~II. Transition probabilities for 36 lines
 of Sn~II arising from the $5s^2ns$, $5s^2np$, $5s^2nd$, $5s^2nf$, and
 $5s5p^2$
 configurations of Sn II were evaluated in ~\cite{alonso-00} using
 the Cowan code.
Radiative transition probabilities and oscillator strengths for
164 lines arising from the  $5s^2ns$, $5s^2np$, $5s^2nd$,
$5s^2nf$, $5s^2ng$, and $5s5p^2$ configurations of Sn~II were
calculated recently by Alonso-Medina {\it et al.\/} in
~\cite{alonso-05}. These values were obtained in intermediate
coupling (IC)  using {\it ab initio} relativistic Hartree-Fock
(HFR) calculations.  The standard method of least square fitting
of experimental energy levels by means of computer codes from
Cowan was used ~\cite{alonso-05} to calculate IC transition rates.
 Recently, energies of the $5s^25p_j$, $5s5p^2$,
$5s^26s_{1/2}$, $5s^25d_j$, and $5s^26p_j$ states in Sn~II were
evaluated by Dzuba and Flambaum in \cite{dzuba-05} using many-body
perturbation theory  (MBPT).  It was underlined that correlations
and relativistic corrections were important. The screening of the
Coulomb interaction and hole-particle interaction was included in
all orders of the MBPT \cite{dzuba-05}.

The experimental  study of atomic lifetimes in gallium, indium,
and thallium was carried   by Andersen and S{\o}rensen
~\cite{life-73} using beam-foil technique. Results for the
$5s^26s$ and $5s^2nd$ ($n$ = 5--7) levels in In~I were given in
~\cite{life-73}. Lifetimes of the $5s^2ns$ and $5s^2nd$ ($n\leq$ =
20) states in indium measured using pulsed laser excitation of an
atomic beam were reported by J{\"{o}}nsson {\it et al.\/} in
~\cite{life-pra-83}. Determination of radiative lifetimes of the
$5s^26p$,  $5s^2ns$, and $5s^2nd$ ($n\leq$ = 10) levels in In~I
using a pulsed laser was presented in
Refs.~\cite{life-astr-83,life-jpb-83}. The atoms were excited in
an atomic beam, with a nitrogen-laser-pumped dye laser. The
fluorescence decay from the atoms was observed by a fast
photomultiplier ~\cite{life-astr-83,life-jpb-83}. The optical
emission  from a laser produced plasma generated by 1064~nm
irradiation of Sn/Pb alloy targets at a flux of
2$\times$10$^{10}$~Wcm$^{-2}$ was recorded and analyzed between
200 and 700~nm \cite{alonso-00}. Experimental transition
probabilities for 36 lines of Sn~II arising from the $5s^2ns$,
$5s^2np$, $5s^2nd$, $5s^2nf$, and
 $5s5p^2$ configurations of Sn~II were determined by
Alonso-Medina {\it et al.\/} in ~\cite{alonso-00}. Lifetime
measurements for levels arising from  the $5s^25d$ and $5s^24f$
configurations in Sn~II were presented by Schectman {\it et al.\/}
in ~\cite{life-00}. These measurements utilized the University of
Toledo Heavy Ion Accelerator Beam Foil Facility. The results were
discussed in the context of interpreting vacuum ultraviolet
absorption spectra observed with the Goddard High Resolution
Spectrograph on board the Hubble Space Telescope ~\cite{life-00}.

Hyperfine structure of the $5s^25p_j$ states of In$^{115}$ and
In$^{113}$ were measured by the magnetic-resonance method
~\cite{hyp-57a,hyp-57b}. An atomic beam irradiated by a narrow
band dye laser was used in \cite{hyp-5s} to observe resonance
fluorescence in free indium atom. From the resonance frequencies,
values for the hyperfine structure of the $5s^26s$ level in
In$^{115}$ and In$^{113}$ were derived by Neijzen and
D{\"{o}}nszelmann in ~\cite{hyp-5s}. The spin-forbidden  $5s^25p\
^2P$ –- $5s5p^2\ ^4P$ In$^{115}$ transition was analyzed
 and absolute wavelengths, hyperfine constants A and
B, as well as  improved energy level values were reported  by
Karlsson and
 Litz\'{e}n in ~\cite{hyp-01}.

A high-resolution study of the  $\lambda$ = 451.1~nm transition in
In~I using CW dye laser was reported by Zaal {\it et al.\/} in
~\cite{laser-78}. The blue dye laser set-up was tested on the
$5s^25p_{3/2}$--$5s^26s_{1/2}$  $\lambda$ = 451.1~nm transition in
natural indium ~\cite{laser-78}. Proposal for laser on
self-terminating transition in blue spectral range on indium atom
transition at 451.1~nm was presented recently by Riyves {\it et
al.\/} in ~\cite{laser-03}. The spectroscopy of dense In vapor was
studied recently via resonant pulsed laser excitation at $\lambda$
= 410.13~nm  (the $5s^25p_{1/2}$--$5s^26s_{1/2}$ transition)
~\cite{laser-07}.

In this paper, we conduct both relativistic many-body perturbation
theory (RMBPT) and all-order single-double (SD) calculations of In~I
and Sn~II properties. Such calculations permit one to investigate
convergence of perturbation theory and estimate the uncertainty of
theoretical predictions. We evaluate reduced matrix elements,
oscillator strengths, and transition rates for possible
$5s^2nl-5s^2n'l'$ electric-dipole transitions in In~I and Sn~II and
calculate the lifetimes of the corresponding levels. Our results are
compared with theoretical results from
Refs.~\cite{migd76,gruzdev78,migd79,migd94,alonso-00,alonso-05} and
with measurements from Refs.
\cite{life-73,life-pra-83,life-jpb-83,life-astr-83} in In~I and
Refs.~\cite{alonso-00,life-00} in Sn~II.  We also calculate
hyperfine constants $A$ for the $5s^2np_j$ $(n= 5-8)$,
$5s^2ns_{1/2}$ $(n= 6-9)$, and $5s^2nd_j$ $(n= 5-8)$ states in
$^{115}$In using the relativistic MBPT and SD all-order methods.
Where possible, we compare our results with the measurements from
Refs.~\cite{hyp-5s,hyp-57a,hyp-57b}.

 We consider
the three-electron system [Ni]$4s^24p^64d^{10}5s^2nl$
  in In~I and Sn~II as a one-electron
$nl$ system with  [Ni]$4s^24p^64d^{10}5s^2$ core.
 Recently, the relativistic all-order  method was used to evaluate
the excitation energies, oscillator strengths, transition rates,
and lifetimes in Ga~I \cite{ga-06} as well as   in Tl~I and
Tl-like Pb \cite{tl-05}. The [Ni]$4s^2nl$ states in Ga~I were
treated in \cite{ga-06} as the $nl$ one-electron system with
[Ni]$4s^2$ core and [Xe]$4f^{14}5d^{10}6s^2nl$ states in Tl~I and
Pb~II were evaluated in \cite{tl-05} as  the $nl$ one-electron
system with [Xe]$4f^{14}5d^{10}6s^2$ core.

To summarize, this work presents both a systematic calculation of
various properties of In~I and Sn~II, and a study of the
importance of the high-order correlation corrections to these
properties. We conclude that all-order SD method, in general,
produce more  accurate values than the third order MBPT and can be
used for the accurate calculation of In and Sn$^+$ properties. By
comparing the all-order and third-order MBPT results, we were able
to study the relative importance of the correlation corrections
for different properties and single out the cases where the
treatment of In as a three-particle system may be important, i.e.
the cases where significant discrepancies between theory and
experiment persist even for the all-order calculations. The
development of the all-order approach  that is capable to fully
treat In or Sn$^+$ as a three-particle system is a difficult
problem~\cite{DFK96a,DFK96b,Koz04} and the initial studies  of the
applicability of the all-order method to such systems may be
useful. We find that  the all-order SD method works relatively
well for In~I even without explicit consideration of the
three-particle states. For Sn~II, the convergence of MBPT
expansion is worse than for In, particularly for $d$-wave, where
SD equations diverge. That is caused by the strong interaction
between $5s^2 nd$ configurations  and low-lying $5s5p^2$
configuration, which corresponds to the excitation from the core.

In the next section, we briefly review  the RMBPT theory and
all-order SD method for the calculation of atomic properties  of
the atoms with one unpaired electron. The energies are given in
Table~\ref{tab-esd}. Extension of the theory to one-electron
matrix elements is discussed in Sec.~III. Our results for E1
transitions are listed in Tables~\ref{tab-osc1}
--~\ref{tab-tran-com-50}. Calculated and experimental lifetimes
for In are given in Table~\ref{tab-life}. Magnetic hyperfine structure
of In is discussed in Sec.~IV and results are summarized  in
Table~\ref{tab-hyp}.

\section{Energies of  I\lowercase{n}~I  and S\lowercase{n}~II }

\begin{table*}[tp]
\caption{\label{tab-esd} Valence energies in different
approximations for In~I and Sn~II in cm$^{-1}$. We calculate
zeroth-order (DF), single-double Coulomb correction
$E^\text{{SD}}$, and the part of third-order
$E^{(3)}_\text{{extra}}$  which is not included in the
$E^\text{{SD}}$. Breit corrections $B^{(n)}$ are calculated in
first and second orders. The sum of these five terms $E_{\rm
tot}^{\rm SD}$ is compared with experimental energies
$E_\text{{NIST}}$~\protect\cite{nist}, $\delta E^{\rm SD} = E_{\rm
tot}^{\rm SD} - E_\text{{NIST}}$. The differences $\delta E^{(2)}$
and $\delta E^{(3)}$ between    total energies ($E^{(2)}_\text{
tot} = E^{(0)} + E^{(2)} + B^{(1)} + B^{(2)}$, $E^{(3)}_\text{
tot} = E^{(2)}_\text{ tot}+ E^{(3)}$) and experimental energies
$E_\text{{NIST}}$~\protect\cite{nist} are given for comparison.  }
\begin{ruledtabular}
\begin{tabular}{lrrrrrrrrrr}
\multicolumn{1}{c}{$nlj$ } & \multicolumn{1}{c}{$E^\text{DF}$} &
\multicolumn{1}{c}{$E^\text{SD}$} &
\multicolumn{1}{c}{$E^{(3)}_\text{{extra}}$} &
\multicolumn{1}{c}{$B^{(1)}$} & \multicolumn{1}{c}{$B^{(2)}$} &
\multicolumn{1}{c}{$E_\text{ tot}^{\rm SD}$} &
\multicolumn{1}{c}{$E_\text{{NIST}}$} & \multicolumn{1}{c}{$\delta
E^{(2)}$} & \multicolumn{1}{c}{$\delta E^{(3)}$}&
\multicolumn{1}{c}{$\delta E^{\rm SD}$}
\\
\hline
  \multicolumn{11}{c}{In~I}\\
  $5p_{1/2}$& -41507&  -5554&    913& 105& -146& -46189& -46670&-2163&  547&  481\\
  $5p_{3/2}$& -39506&  -5378&    912&  73& -132& -44031& -44457&-2040&  557&  426\\
  $5d_{3/2}$& -12390&  -1350&    161&   2&   -4& -13581& -13778&  256&  493&  197\\
  $5d_{5/2}$& -12374&  -1337&    160&   1&   -4& -13554& -13755&  260&  493&  201\\
  $6s_{1/2}$& -20572&  -2096&    232&  12&  -18& -22442& -22297& -237&  249& -145\\
  $6p_{1/2}$& -13979&   -964&    113&  13&  -16& -14833& -14853& -150&  122&   20\\
  $6p_{3/2}$& -13719&   -919&    112&  10&  -16& -14532& -14555& -155&  115&   23\\
  $6d_{3/2}$&  -6955&   -558&     69&   1&   -2&  -7445&  -7809&  375&  479&  364\\
  $6d_{5/2}$&  -6946&   -554&     68&   1&   -2&  -7433&  -7697&  277&  379&  264\\
  $7s_{1/2}$&  -9867&   -584&     72&   4&   -6& -10381& -10368&  -95&   64&  -13\\
  $7p_{1/2}$&  -7488&   -349&     41&   5&   -6&  -7797&  -7809&  -56&   45&   12\\
  $7p_{3/2}$&  -7388&   -335&     41&   4&   -6&  -7684&  -7697&  -58&   42&   13\\
  $8s_{1/2}$&  -5816&   -251&     32&   2&   -3&  -6036&  -6033&  -45&   27&   -3\\
  $4f_{5/2}$&  -6863&   -118&     14&   0&    0&  -6967&  -6963&   -4&    9&   -4\\
  $4f_{7/2}$&  -6863&   -118&     14&   0&    0&  -6967&  -6962&   -5&    8&   -5\\
  $5f_{5/2}$&  -4393&    -67&      8&   0&    0&  -4452&  -4450&   -3&    5&   -2\\
  $5f_{7/2}$&  -4393&    -67&      8&   0&    0&  -4452&  -4450&   -3&    5&   -2\\
  $7d_{3/2}$&  -4441&   -285&     35&   0&   -1&  -4692&  -4834&  148&  201&  142\\
  $7d_{5/2}$&  -4436&   -283&     35&   0&   -1&  -4685&  -4808&  129&  181&  123\\
  $8p_{1/2}$&  -4687&   -168&     20&   2&   -3&  -4836&  -4843&  -28&   21&    7\\
  $8p_{3/2}$&  -4638&   -162&     20&   2&   -3&  -4781&  -4789&  -28&   21&    8\\
  $8d_{3/2}$&  -3078&   -171&     21&   0&   -1&  -3229&  -3334&  112&  144&  105\\
  $8d_{5/2}$&  -3075&   -170&     21&   0&   -1&  -3225&  -3315&   98&  129&   90\\
  $9s_{1/2}$&  -3837&   -132&     17&   1&    0&  -3951&  -3951&  -24&   15&    0\\
\hline
  \multicolumn{11}{c}{Sn~II}\\
  $5p_{1/2}$&-111452&  -6848&  1014&  206& -233&-117313&-118017&-2578&  712&  704\\
  $5p_{3/2}$&-107358&  -6719&  1032&  146& -216&-113115&-113766&-2459&  736&  651\\
  $6s_{1/2}$& -57995&  -3597&   467&   35&  -44& -61134& -61131& -641&  435&   -3\\
  $6p_{1/2}$& -44483&  -2244&   271&   39&  -38& -46455& -46523& -316&  314&   68\\
  $6p_{3/2}$& -43691&  -2133&   258&   28&  -37& -45575& -45640& -297&  300&   65\\
  $7s_{1/2}$& -30735&  -1230&   172&   14&  -17& -31796& -31737& -343&   55&  -59\\
  $4f_{5/2}$& -27689&  -1189&   147&    0&   -1& -28732& -28731&  -24&  206&   -1\\
  $4f_{7/2}$& -27691&  -1193&   147&    0&   -1& -28738& -28725&  -34&  196&  -13\\
  $7p_{1/2}$& -25253&   -930&   115&   16&  -16& -26068& -26114& -107&  153&   46\\
  $7p_{3/2}$& -24917&   -895&   109&   12&  -15& -25706& -25751&  -94&  152&   45\\
  $8s_{1/2}$& -19133&   -629&    84&    7&   -9& -19680& -19615& -162&   26&  -65\\
  $5f_{5/2}$& -17759&   -670&    85&    0&   -1& -18345& -18358&  -13&  131&   13\\
  $5f_{7/2}$& -17761&   -674&    86&    0&   -1& -18350& -18352&  -22&  121&    2\\
  $8p_{1/2}$& -16354&   -487&    60&    8&   -8& -16781& -16821&  -31&  102&   40\\
  $8p_{3/2}$& -16179&   -472&    57&    6&   -8& -16596& -16630&  -29&   97&   34\\
  $9s_{1/2}$& -13070&   -301&    47&    4&    0& -13320& -13337&  -88&   13&   17\\
\end{tabular}
\end{ruledtabular}
\end{table*}

We  start from the ``no-pair'' Hamiltonian  \cite{sucher} in
the second quantization form
\begin{align}
\label{eq1}
H &=H_{0}+V_{I}\,,\\
\label{eq2}
H_{0} &=\sum_{i}\varepsilon _{i}a_{i}^{+}a_{i}\,,\\
\label{eq3}
V_{I} &=\sum_{ijkl}g_{ijkl}a_{i}^{+}a_{j}^{+}a_{l}a_{k}\, ,
\end{align}
where negative energy (positron) states are excluded from the sums;
$\varepsilon _{i}$ are eigenvalues of the one-electron DF equations
with a frozen core, and $g_{ijkl}$ is the Coulomb two-particle
matrix element.

Considering  neutral In as a one-electron system we use $V^{N-1}$ DF
potential [Ni]$4s^{2}4p^{6}4d^{10}5s^{2}$ to calculate DF orbitals
and energies $\varepsilon _{i}$. There are a number of advantages
associated with this potential, including a greatly reduced number
of the Goldstone diagrams \cite{goldstone}, which leads to important
simplifications in calculation. For example, when considering the
total energy of different valence states of a one-electron atom,
that energy can be written as
\begin{equation}\label{eq4}
\ E=E_{v}+E_\text{core}\,,
\end{equation}
where $E_\text{core}$ is the same for all valence states $v$. The
first-order correlation correction  to valence removal energies
vanishes for a $V^{N-1}$ DF potential and the first nonvanishing
corrections appear in the second order~\cite{wrj-en}:
\begin{align}
E_{v}^{(2)}
&=\sum_{mn}\sum_{a}\frac{g_{avmn}(g_{mnav}-g_{mnva})}{\varepsilon
_{a}+\varepsilon _{v}-\varepsilon _{n}-\varepsilon _{m}}
\nonumber\\
\label{eq5}
&+\sum_{n}\sum_{ab}%
\frac{g_{nvba}(g_{abnv}-g_{abvn})}{\varepsilon _{a}+\varepsilon
_{b}-\varepsilon _{n}-\varepsilon _{v}}\,.
\end{align}
We use indexes  $a$ and $b$ to label  core states and $m$ and $n$
to designate any excited  states. The second-order Coulomb-Breit
contribution $B_{v}^{(2)}$ is obtained from the $E_{v}^{(2)}$
expression (\ref{eq5}) by changing $g_{ijkl}\rightarrow
g_{ijkl}+b_{ijkl}$ and keeping only terms that are linear in
$b_{ijkl}$ that is a two-particle matrix element of the Breit
interaction \cite{li-en}:
\begin{equation}\label{eq-br}
B=-\frac{\alpha}{r_{12}}\left[ \bm{\alpha}_1\bm{\alpha}_2-%
\tfrac{1}{2}\left[ \bm{\alpha}_{1} \bm{\alpha}_{2}-\left(
\bm{\alpha}_{1}\widehat{{\bf r}}_{12}\right) \left( \bm{\alpha }%
_{2}\widehat{{\bf r}}_{12}\right) \right] \right],
\end{equation}
where $\bm{\alpha}_{1}$ is the Dirac matrix, $\widehat{{\bf r}}_{12}={\bf r}%
_{12}/r_{12}$, and $\alpha $ is the fine structure constant.  The
first-order Breit correction is $B_{v}^{(1)}=\sum_{a}\left[
b_{vava}-b_{vaav}\right] =-\sum_{a}b_{vaav}$, where direct term
vanishes after summing over closed shells.

Even though the number of Goldstone diagrams for the $V^{N-1}$ DF
potential is much smaller than in general case, the third-order
expression for energy correction still includes 52 terms. The
corresponding formula for $E^{(3)}_v$ was presented by Blundell {\it
et al.} in Ref.~\cite{tl-en}, where 52 terms were combined into 12
groups with distinct energy denominators:
\begin{equation}\label{eq7}
E_{v}^{(3)}=E_{A}^{(3)} +\cdots +E_{L}^{(3)}\,.
\end{equation}
Expression (\ref{eq7}) includes terms with one-, two-, three-, and
four-particle sums over virtual states in addition to sums over core
states.

The all-order SD method was discussed previously in
Refs.~\cite{blundell-li,Liu,blundell-cs,safr-na,safr-alk,gold,tl-05}.
Briefly, we represent the wave function $\Psi _{v}$ of the atom with
one valence electron as $\Psi _{v}\cong \Psi _{v}^{\rm SD}$:
\begin{eqnarray}\label{eq8}
\Psi _{v}^{\rm SD}\! &=&\!\left[\! 1+\sum_{ma}\rho _{ma}a_{m}^{+}a_{a}+\tfrac{1}{2}%
\sum_{mnab}\rho _{mnab}a_{m}^{+}a_{n}^{+}a_{b}a_{a}\right.
\\
\nonumber &+&\left.\sum_{m\neq v}\rho
_{mv}a_{m}^{+}a_{v}+\frac{1}{2}\sum_{mna}\rho
_{mnva}a_{m}^{+}a_{n}^{+}a_{a}a_{v}\right] \Phi _{v}\,,
\end{eqnarray}
where $\Phi _{v}$ is the lowest-order atomic wave function, which
is taken to be the frozen-core DF wave function of a state $v$.
The coupled equations for the single- ($\rho _{mv}$ and $\rho
_{ma}$) and double-excitation coefficients $\rho _{mnva}$ and
$\rho _{mnab}$ are obtained by substituting the wave function
$\Psi _{v}^{\rm SD}$ into the many-body Schr\"{o}dinger equation,
with Hamiltonian given by Eqs.~(\ref{eq1}--\ref{eq3}). Note that
we again start from $V^{n-1}$ DF potential.
 The coupled
equations  for the excitation coefficients are solved iteratively.
In the following sections, the resulting excitation coefficients
are used to evaluate hyperfine constants and transition matrix
elements.

The valence energy $E_{v}^{\rm SD}$  is given by
\begin{equation}\label{eq9}
E_{v}^{\rm SD}=\sum_{ma}\widetilde{g}_{vavm}\rho _{ma}+\sum_{mab}g_{abvm}%
\widetilde{\rho }_{mvab}+\sum_{mna}g_{vamn}\widetilde{\rho
}_{mnva}\,.
\end{equation}
This expression does not include a certain part of the third-order
MBPT contribution. This part of the third-order contribution,
$E_{v,\rm extra}^{(3)}$, is given in Ref.~\cite{safr-na}  and has
to be calculated separately. We use our third-order energy code to
separate out $E_{v,\rm extra}^{(3)}$ and add it to the $E_{v}^{\rm
SD}$. We drop the index $v$ in the $E_{v}^{(2)}$, $E_{v}^{(3)}$,
and $E_{v}^{\rm SD}$ designations in the text and tables below.

We use B-splines \cite{Bspline} to generate a basis set of DF wave
functions for the calculations of MBPT and all-order expressions.
Typically, we use 40 or 50 splines of order $k$ = 7 or 9,
respectively, for each partial wave (see below for more details).
Basis orbitals for In~I and Sn~II are constrained to cavities of
radii $R=95$ a.u.\ and $R=85$ a.u.\, respectively. The cavity
radii are chosen large enough to accommodate all orbitals
considered in this paper and small enough for 50 splines to
approximate inner-shell DF wave functions with good precision.

Results of our all-order SD calculations of energies for the
lowest states of neutral In and In-like Sn ion are given in
Table~\ref{tab-esd}. Our final answer $E^\text{{SD}}_{\rm tot}$
also includes the part of the third-order energies omitted in the
SD calculation $E^{(3)}_\text{{extra}}$, as well as the
first-order Breit correction $B^{(1)}$ and the second-order
Coulomb-Breit $B^{(2)}$ correction. Theoretical values are
compared with the recommended values $E_\mathrm{NIST}$ from the
National Institute of Standards and Technology database
\cite{nist}, $\delta E^\text{SD}=E^\text{{SD}}_{\rm
tot}-E_\mathrm{NIST}$. For comparison, we also give the
differences between   the second-order and third-order MBPT
calculations and experimental values in columns labelled  $\delta
E^{(2)}$ and $\delta E^{(3)}$. In Sn~II the all-order SD equations
for $d$-wave do not converge and we exclude $d$-orbitals of Sn~II
from Table~\ref{tab-esd}.

\begin{table*}[p]
\caption{\label{tab-osc1} Wavelengths $\lambda$ (\AA), transition
rates $A_r$ (s$^{-1}$), oscillator strengths ($f$), and line
strengths $S$ (a.u.) for transitions in In~I calculated in all-order
perturbation theory. Numbers in brackets represent powers of 10. }
\begin{ruledtabular}
\begin{tabular}{llrlllllrlll}
\multicolumn{2}{c}{Transition} &
\multicolumn{1}{c}{$\lambda$} &
\multicolumn{1}{c}{$A_r$} &
\multicolumn{1}{c}{$f$} &
\multicolumn{1}{c}{$S$} &
\multicolumn{2}{c}{Transition} &
\multicolumn{1}{c}{$\lambda$} &
\multicolumn{1}{c}{$A_r$} &
\multicolumn{1}{c}{$f$} &
\multicolumn{1}{c}{$S$} \\
\hline
$5p_{1/2}$&$ 6s_{1/2}$&     4153&  5.15[7]&  1.33[-1]&  3.64[ 0]&$4f_{5/2}$&$ 8d_{5/2}$&    26532&  3.32[5]&  3.51[-2]&  1.84[ 1]\\
$5p_{1/2}$&$ 5d_{3/2}$&     3045&  1.30[8]&  3.61[-1]&  7.24[ 0]&$4f_{7/2}$&$ 7d_{5/2}$&    42937&  1.65[5]&  3.42[-2]&  3.86[ 1]\\
$5p_{1/2}$&$ 7s_{1/2}$&     2792&  1.37[7]&  1.60[-2]&  2.93[-1]&$4f_{7/2}$&$ 8d_{5/2}$&    26532&  6.65[4]&  5.26[-3]&  3.68[ 0]\\
$5p_{1/2}$&$ 6d_{3/2}$&     2578&  3.63[7]&  7.24[-2]&  1.23[ 0]&$8p_{1/2}$&$ 7d_{3/2}$&   529101&  4.67[3]&  3.92[-1]&  1.37[ 3]\\
$5p_{1/2}$&$ 8s_{1/2}$&     2493&  5.81[6]&  5.41[-3]&  8.88[-2]&$8p_{1/2}$&$ 8d_{3/2}$&    61275&  3.79[5]&  4.27[-1]&  1.72[ 2]\\
$5p_{1/2}$&$ 7d_{3/2}$&     2410&  1.47[7]&  2.56[-2]&  4.06[-1]&$8p_{3/2}$&$ 7d_{3/2}$&   740741&  3.40[2]&  2.80[-2]&  2.73[ 2]\\
$5p_{1/2}$&$ 8d_{3/2}$&     2329&  7.49[6]&  1.22[-2]&  1.87[-1]&$8p_{3/2}$&$ 7d_{5/2}$&   704225&  2.38[3]&  2.66[-1]&  2.47[ 3]\\
$5p_{3/2}$&$ 6s_{1/2}$&     4576&  9.05[7]&  1.42[-1]&  8.56[ 0]&$8p_{3/2}$&$ 8d_{3/2}$&    63371&  8.38[4]&  5.05[-2]&  4.21[ 1]\\
$5p_{3/2}$&$ 5d_{3/2}$&     3266&  2.49[7]&  3.98[-2]&  1.71[ 0]&$8p_{3/2}$&$ 8d_{5/2}$&    63211&  4.95[5]&  4.44[-1]&  3.70[ 2]\\
$5p_{3/2}$&$ 5d_{5/2}$&     3264&  1.47[8]&  3.53[-1]&  1.52[ 1]&$5f_{5/2}$&$ 8d_{3/2}$&    79681&  1.22[5]&  7.75[-2]&  1.22[ 2]\\
$5p_{3/2}$&$ 7s_{1/2}$&     2977&  2.31[7]&  1.53[-2]&  6.02[-1]&$5f_{5/2}$&$ 8d_{5/2}$&    79428&  5.70[3]&  5.39[-3]&  8.46[ 0]\\
$5p_{3/2}$&$ 6d_{3/2}$&     2734&  6.88[6]&  7.71[-3]&  2.78[-1]&$5f_{7/2}$&$ 8d_{5/2}$&    79428&  1.14[5]&  8.09[-2]&  1.69[ 2]\\
$5p_{3/2}$&$ 6d_{5/2}$&     2734&  4.05[7]&  6.81[-2]&  2.45[ 0]&$6s_{1/2}$&$ 6p_{1/2}$&    13669&  1.43[7]&  4.02[-1]&  3.61[ 1]\\
$5p_{3/2}$&$ 8s_{1/2}$&     2639&  9.72[6]&  5.07[-3]&  1.76[-1]&$6s_{1/2}$&$ 6p_{3/2}$&    13146&  1.57[7]&  8.13[-1]&  7.03[ 1]\\
$5p_{3/2}$&$ 7d_{3/2}$&     2547&  2.79[6]&  2.71[-3]&  9.08[-2]&$6s_{1/2}$&$ 7p_{1/2}$&     7002&  1.40[6]&  1.03[-2]&  4.75[-1]\\
$5p_{3/2}$&$ 7d_{5/2}$&     2546&  1.63[7]&  2.38[-2]&  7.97[-1]&$6s_{1/2}$&$ 7p_{3/2}$&     6949&  1.96[6]&  2.84[-2]&  1.30[ 0]\\
$5p_{3/2}$&$ 8d_{3/2}$&     2456&  1.43[6]&  1.29[-3]&  4.18[-2]&$6s_{1/2}$&$ 8p_{1/2}$&     5806&  4.07[5]&  2.06[-3]&  7.86[-2]\\
$5p_{3/2}$&$ 8d_{5/2}$&     2456&  8.32[6]&  1.13[-2]&  3.65[-1]&$6s_{1/2}$&$ 8p_{3/2}$&     5787&  6.40[5]&  6.42[-3]&  2.45[-1]\\
$6p_{1/2}$&$ 5d_{3/2}$&    69156&  1.57[5]&  2.25[-1]&  1.03[ 2]&$5d_{3/2}$&$ 7p_{1/2}$&    18116&  7.58[5]&  1.86[-2]&  4.45[ 0]\\
$6p_{1/2}$&$ 7s_{1/2}$&    22594&  3.48[6]&  2.66[-1]&  3.96[ 1]&$5d_{3/2}$&$ 7p_{3/2}$&    17765&  6.13[4]&  2.90[-3]&  6.78[-1]\\
$6p_{1/2}$&$ 6d_{3/2}$&    13512&  7.35[6]&  4.02[-1]&  3.58[ 1]&$5d_{3/2}$&$ 4f_{5/2}$&    15798&  1.32[7]&  7.43[-1]&  1.55[ 2]\\
$6p_{1/2}$&$ 8s_{1/2}$&    11463&  1.13[6]&  2.22[-2]&  1.68[ 0]&$5d_{3/2}$&$ 8p_{1/2}$&    11816&  3.07[5]&  3.21[-3]&  5.00[-1]\\
$6p_{1/2}$&$ 7d_{3/2}$&     9903&  3.79[6]&  1.12[-1]&  7.27[ 0]&$5d_{3/2}$&$ 8p_{3/2}$&    11741&  2.46[4]&  5.09[-4]&  7.87[-2]\\
$6p_{1/2}$&$ 8d_{3/2}$&     8665&  2.19[6]&  4.92[-2]&  2.81[ 0]&$5d_{3/2}$&$ 5f_{5/2}$&    11312&  5.54[6]&  1.59[-1]&  2.37[ 1]\\
$6p_{3/2}$&$ 5d_{3/2}$&    86580&  1.60[4]&  1.80[-2]&  2.05[ 1]&$5d_{5/2}$&$ 7p_{3/2}$&    17838&  5.67[5]&  1.80[-2]&  6.36[ 0]\\
$6p_{3/2}$&$ 5d_{5/2}$&    84890&  1.03[5]&  1.66[-1]&  1.86[ 2]&$5d_{5/2}$&$ 4f_{5/2}$&    15855&  9.46[5]&  3.56[-2]&  1.12[ 1]\\
$6p_{3/2}$&$ 7s_{1/2}$&    24184&  6.36[6]&  2.79[-1]&  8.88[ 1]&$5d_{5/2}$&$ 4f_{7/2}$&    15855&  1.42[7]&  7.13[-1]&  2.23[ 2]\\
$6p_{3/2}$&$ 6d_{3/2}$&    14065&  1.54[6]&  4.56[-2]&  8.45[ 0]&$5d_{5/2}$&$ 8p_{3/2}$&    11773&  2.29[5]&  3.17[-3]&  7.36[-1]\\
$6p_{3/2}$&$ 6d_{5/2}$&    14043&  9.12[6]&  4.04[-1]&  7.48[ 1]&$5d_{5/2}$&$ 5f_{5/2}$&    11342&  3.93[5]&  7.59[-3]&  1.70[ 0]\\
$6p_{3/2}$&$ 8s_{1/2}$&    11858&  1.96[6]&  2.06[-2]&  3.22[ 0]&$5d_{5/2}$&$ 5f_{7/2}$&    11342&  5.90[6]&  1.52[-1]&  3.40[ 1]\\
$6p_{3/2}$&$ 7d_{3/2}$&    10197&  7.61[5]&  1.19[-2]&  1.59[ 0]&$7s_{1/2}$&$ 7p_{1/2}$&    39370&  2.42[6]&  5.62[-1]&  1.46[ 2]\\
$6p_{3/2}$&$ 7d_{5/2}$&    10190&  4.54[6]&  1.06[-1]&  1.42[ 1]&$7s_{1/2}$&$ 7p_{3/2}$&    37750&  2.63[6]&  1.12[ 0]&  2.79[ 2]\\
$6p_{3/2}$&$ 8d_{3/2}$&     8889&  4.31[5]&  5.11[-3]&  5.98[-1]&$7s_{1/2}$&$ 8p_{1/2}$&    18238&  3.95[5]&  1.97[-2]&  2.37[ 0]\\
$6p_{3/2}$&$ 8d_{5/2}$&     8886&  2.58[6]&  4.58[-2]&  5.36[ 0]&$7s_{1/2}$&$ 8p_{3/2}$&    18060&  5.16[5]&  5.04[-2]&  6.00[ 0]\\
$7p_{1/2}$&$ 6d_{3/2}$&   229885&  2.00[4]&  3.17[-1]&  4.80[ 2]&$6d_{3/2}$&$ 4f_{5/2}$&   266667&  1.11[4]&  1.77[-1]&  6.23[ 2]\\
$7p_{1/2}$&$ 8s_{1/2}$&    56883&  8.36[5]&  4.06[-1]&  1.52[ 2]&$6d_{3/2}$&$ 8p_{1/2}$&    39872&  3.44[5]&  4.10[-2]&  2.15[ 1]\\
$7p_{1/2}$&$ 7d_{3/2}$&    31928&  1.33[6]&  4.05[-1]&  8.51[ 1]&$6d_{3/2}$&$ 8p_{3/2}$&    39032&  2.82[4]&  6.44[-3]&  3.31[ 0]\\
$7p_{1/2}$&$ 8d_{3/2}$&    21858&  8.39[5]&  1.20[-1]&  1.73[ 1]&$6d_{3/2}$&$ 5f_{5/2}$&    34662&  1.94[6]&  5.24[-1]&  2.39[ 2]\\
$7p_{3/2}$&$ 6d_{3/2}$&   306748&  1.68[3]&  2.37[-2]&  9.59[ 1]&$6d_{5/2}$&$ 4f_{5/2}$&   274725&  7.24[2]&  8.19[-3]&  4.45[ 1]\\
$7p_{3/2}$&$ 6d_{5/2}$&   296736&  1.12[4]&  2.22[-1]&  8.67[ 2]&$6d_{5/2}$&$ 4f_{7/2}$&   274725&  1.09[4]&  1.64[-1]&  8.89[ 2]\\
$7p_{3/2}$&$ 8s_{1/2}$&    60643&  1.53[6]&  4.20[-1]&  3.36[ 2]&$6d_{5/2}$&$ 8p_{3/2}$&    39200&  2.61[5]&  4.01[-2]&  3.10[ 1]\\
$7p_{3/2}$&$ 7d_{3/2}$&    33080&  2.87[5]&  4.72[-2]&  2.05[ 1]&$6d_{5/2}$&$ 5f_{5/2}$&    34795&  1.40[5]&  2.54[-2]&  1.74[ 1]\\
$7p_{3/2}$&$ 7d_{5/2}$&    33003&  1.70[6]&  4.16[-1]&  1.81[ 2]&$6d_{5/2}$&$ 5f_{7/2}$&    34795&  2.10[6]&  5.07[-1]&  3.49[ 2]\\
$7p_{3/2}$&$ 8d_{3/2}$&    22391&  1.75[5]&  1.31[-2]&  3.87[ 0]&$8s_{1/2}$&$ 8p_{1/2}$&    84388&  6.59[5]&  7.04[-1]&  3.91[ 2]\\
$7p_{3/2}$&$ 8d_{5/2}$&    22371&  1.04[6]&  1.17[-1]&  3.44[ 1]&$8s_{1/2}$&$ 8p_{3/2}$&    80710&  7.18[5]&  1.40[ 0]&  7.45[ 2]\\
$4f_{5/2}$&$ 7d_{3/2}$&    43066&  1.77[5]&  3.28[-2]&  2.79[ 1]&$7d_{3/2}$&$ 5f_{5/2}$&   531915&  4.89[3]&  3.11[-1]&  2.18[ 3]\\
$4f_{5/2}$&$ 7d_{5/2}$&    42937&  8.24[3]&  2.28[-3]&  1.93[ 0]&$7d_{5/2}$&$ 5f_{5/2}$&   552486&  3.12[2]&  1.43[-2]&  1.56[ 2]\\
$4f_{5/2}$&$ 8d_{3/2}$&    26560&  7.12[4]&  5.02[-3]&  2.63[ 0]&$7d_{5/2}$&$ 5f_{7/2}$&   552486&  4.68[3]&  2.85[-1]&  3.11[ 3]\\
\end{tabular}
\end{ruledtabular}
\end{table*}

The largest correlation contribution to the valence energy comes
from the second-order term, $E^{(2)}$. As we have discussed
above, this term is simple  to calculate in comparison with
$E^{(3)}$ and $E^{\rm SD}$ terms. Thus, we calculate $E^{(2)}$
with better accuracy than $E^{(3)}$ and $E^{\rm SD}$. To increase
the accuracy of the $E^{(2)}$ calculations, we use 50 splines of
order $k=9$ for each partial wave and include partial waves up to
$l_{\text{max}}=10$. Then, the final value is extrapolated to
account for contributions from higher partial waves (see, for
example, Refs.~\cite{be-en,be3-en}).We estimate the numerical
uncertainty of $E^{(2)}$ caused by incompleteness of the basis set
to be approximately 10~cm$^{-1}$ or less, depending on the valence
state.

Owing to the numerical complexity of the $E^{\rm SD}$ calculation,
we use $l_{\text{max}}$ = 6  and 40 splines of order $k=7$. As we
noted above, the second-order $E^{(2)}$ is included in the $E^{\rm
SD}$ value. Therefore, we use our high-precision calculation of
$E^{(2)}$ described above to account for the contributions of the
higher partial waves by replacement  $E^{(2)}$[$l_{\text{max}}$ =
6] value with the final high-precision second-order value
$E^{(2)}_{\rm final}$:
\[
E_{\rm final}^{\rm SD} = E^{\rm SD} + E_{\rm final}^{(2)} -
E^{(2)}[l_{\rm max} =6]\,.
\]
The size of this correction varies from $\sim 200$~cm$^{-1}$ for the
lowest valence states to $\sim$~1~--~20~cm$^{-1}$ for other valence
states considered in this work.

A lower number of partial waves, $l_{\text{max}}$ =6, is used also
in the third-order calculation. Since the asymptotic
$l$-dependence of the second- and third-order energies are similar
(both fall off as $l^{-4}$), we use the second-order remainder  to
estimate the numerical uncertainties  in the third-order and in
all-order corrections.

In our calculations of the Breit contribution, we use the whole
operator \eqref{eq-br} in the first-order correction $B^{(1)}$,
while the second-order Coulomb-Breit energies $B^{(2)}$ are
evaluated using the unretarded Breit operator, also known as Gaunt
(it is described by the first term in \eqref{eq-br}). Usually
Gaunt part strongly dominates in the Breit corrections to the
valence energies \cite{Gra70}. Table~\ref{tab-esd} shows that
there is strong cancelation between first and second order
corrections. It is in agreement with the well known observation
that Breit interaction for valence electrons is screened by the
core~\cite{LMY89,KPT00a}.

We have also estimated Lamb shift correction to valence energies.
The vacuum-polarization was calculated in the Uehling
approximation. The self-energy contribution is estimated for the
$s$, $p_{1/2}$ and $p_{3/2}$ orbitals by interpolating the values
obtained by \citet{mohr1,mohr2,mohr3} using Coulomb wave
functions. We found, as expected, that Lamb shift correction is
very small ($E_\text{ LS}\leq$ 3~cm$^{-1}$ for In~I  and $E_\text{
LS}\leq$ 10~cm$^{-1}$ for Sn~II ). This is well below the accuracy
of the present theory, and we neglect this contribution in
Table~\ref{tab-esd}.

Comparison of the differences $\delta E^{(2)}=E^{(2)}_\text{
tot}-E_\text{{NIST}}$ and $\delta E^{(3)}=E^{(3)}_\text{
tot}-E_\text{{NIST}}$ given in Table~\ref{tab-esd} shows that
convergence of MBPT series is not very good for both In and
Sn$^+$. In particular, the second-order results for $d$-wave in In
and $f$-wave in Sn$^+$ are even better than the third-order ones.
All-order results are more accurate than the third-order ones, but
the difference is not very large. For $p$-waves, SD calculation
without the third-order correction overestimates valence binding
energies and underestimates it when this correction is included.
For $d$-wave, both variants lead to underestimation of the binding
energy and term $E_{\rm extra}^{(3)}$ worsens the agreement with
the experiment.

We conclude that all-order calculation is generally more accurate
than the third-order MBPT calculation. Account of the missing
third-order terms does not lead to improvement of the accuracy. On
the other hand, this term is generally on the order of  our final
difference with experiment  and can serve as an estimate of the
latter. For most levels, our final accuracy is better than 1\%,
but the accuracy for the $d$-wave of In is noticeably worse. That
can be explained by the existence of the low-lying configuration
$5s5p^2$ which strongly interacts with configurations $5s^2nd$. To
account for this interaction effectively, one needs to consider In
as a three electron atom \cite{kozlov}. The same reason explains
mentioned above divergence of the SD equations for $d$-wave of
Sn~II. Interaction between configurations $5s5p^2$ and $5s^2ns$ is
weaker and SD equations for $s$-wave converge for both atoms
considered here. In the opposite parity there is no such a
low-lying excitation of the shell $5s$, so MBPT works better and
no problems with convergence occur.

In order to study the relative role of the valence correlations we
have performed the second-order RMBPT calculations of atomic
properties of In~I and Sn~II considering these atoms as
three-electron systems. Corresponding variant of RMBPT was developed
in \cite{bor-en,bor-tr,bor3,alum-en,alum-tr}. The energies of the
[He]$2s^22p$, [He]$2s2p^2$, and [He]$2p^3$ states of B-like systems
were presented in Ref.~\cite{bor-en}.   The second-order  RMBPT was
used by Johnson \textit{et al.\/} \cite{3elec} to calculate
[Ne]$3s^23l$ and [Ne]$3p^23s$ states in Al~I and
[Xe]$4f^{14}5d^{10}6s^26pl$ and [Xe]$4f^{14}5d^{10}6s6p^2$ states in
Tl~I. Comparing results obtained for neutral B~I, Al~I, and Tl~I, we
find that the discrepancy between RMBPT and experimental results
increases significantly from B~I to Tl~I. For example, the RMBPT and
NIST values of the $ns^2np\ [^2P_{3/2}-\ ^2P_{1/2}]$
splitting in cm$^{-1}$ for $n=2$ are equal to 17 and 15; for $n=3$
corresponding values are 123 and 112; finaly, for $n=6$ we get 6710
and 7793 respectively. It is evident that for a light system, such
as B~I, the second-order three-electron RMBPT treatment works much
better than for a heavy system, such as Tl~I. For the latter case it
is more appropriate to consider Tl~I as one-electron system with
[Xe]$4f^{14}5d^{10}6s^2$ core but treat correlation more completely.
It was found in ~\cite{tl-05} that in such approach the discrepancy
between the SD and NIST values of the $6s^26p\ [^2P_{3/2}-\
^2P_{1/2}]$ splitting is only 41~cm$^{-1}$ instead of 1083~cm$^{-1}$
obtained  in ~\cite{3elec}. Alternatively, one can use CI+MBPT
method~\cite{DFK96b}, where the discrepancy is 43~cm$^{-1}$~\cite{kozlov}.

The main difference between configurations
[Ni]$4s^24p^64d^{10}5s^2nl$ of In-like ions and [Ne]$3s^2nl$
configurations of Al-like ions is the necessary size of the model
space for valence electrons. For $5l$ electrons in In-like ions we
could not construct sufficiently complete three-electron model space as we did
for $3l$ electrons. Additionally, in In-like ions the $n$ = 4 core
shell is not filled. Obviously, we can not expect the
same accuracy as in the case of Al-like ions
~\cite{alum-en,3elec}.

We tried two model spaces to evaluate energies of In-like ions.
Firstly we constructed the model space including $5s$, $5p$, and
$5d$ electrons, [$spd$] model space. Secondly, the odd-parity model
space was [$5s^25p$ + $5p^3$] and even-parity model space was
[$5s^25d$ + $5s5p^2$]. We found that in the second case the RMBPT
energies were in better agreement  with NIST data \cite{nist}  than
in the case of more complete [$spd$] model space. Theoretical values
of the $5s^25p\ [^2P_{3/2}-\ ^2P_{1/2}]$ splitting were equal to
2669~cm$^{-1}$ and 4889~cm$^{-1}$ in In~I and Sn~II, respectively.
Comparison of these values with the $E^{\rm SD}_{\rm tot}$ values
from Table~\ref{tab-esd} (2158~cm$^{-1}$ in In~I and
4198~cm$^{-1}$ in Sn~II) shows that the one-electron representation
with all-order treatment of correlation correction gives the results
that are in substantially better agreement with experiment than the
three-electron model space theory.
Because of that, we decided not include three-electron results in
the present paper.

\section{Electric-dipole matrix elements, oscillator strengths, transition
rates, and lifetimes in I\lowercase{n}~I and S\lowercase{n}~II}

The one-body matrix element of the operator $Z$ is given by
\cite{blundell-li}:
\begin{equation}\label{eq11}
Z_{wv}=\frac{\left\langle \Psi _{w}\right| Z\left| \Psi _{v}\right\rangle }{%
\sqrt{\left\langle \Psi _{v}|\Psi _{v}\right\rangle \left\langle
\Psi _{w}|\Psi _{w}\right\rangle }}\,,
\end{equation}
where $\Psi _{v,w}$ are exact wave functions for the many-body
``no-pair'' Hamiltonian $H$
\begin{equation}\label{eq12}
H\left| \Psi _{v}\right\rangle =E\left| \Psi _{v}\right\rangle\,.
\end{equation}
In MBPT, we expand the many-electron function $\Psi _{v}$  in powers of
$V_{I}$ as
\begin{equation}\label{eq13}
\left| \Psi _{v}\right\rangle =\left| \Psi _{v}^{(0)}\right\rangle
+\left| \Psi _{v}^{(1)}\right\rangle +\left| \Psi
_{v}^{(2)}\right\rangle +\left| \Psi _{v}^{(3)}\right\rangle +
\cdots\,.
\end{equation}
The denominator in Eq.~(\ref{eq11}) arises from the normalization
condition that starts to contribute in the third order
\cite{equation}. In the lowest order, we find
\begin{equation}\label{eq14}
Z_{wv}^{(1)}=\left\langle \Psi _{w}^{(0)}\right| Z\left| \Psi
_{v}^{(0)}\right\rangle =z_{wv}\,,
\end{equation}
where $z_{wv}$ is the corresponding one-electron matrix element.
Since $\Psi _{w}^{(0)}$  is a DF function  we use $Z^{\rm DF}$
designation instead of $Z^{(1)}$ below.

\begin{table}[hbt]
\caption{\label{tab-osc2} Wavelengths $\lambda$ (\AA), transition
rates $A_r$ (cm$^{-1}$), oscillator strengths ($f$), and line
strengths $S$ (a.u.) for transitions in Sn~II calculated using
all-order  method.}
\begin{ruledtabular}
\begin{tabular}{llrlll}
\multicolumn{2}{c}{Transition} &
\multicolumn{1}{c}{$\lambda$} &
\multicolumn{1}{c}{$A_r$} &
\multicolumn{1}{c}{$f$} &
\multicolumn{1}{c}{$S$} \\
\hline
$5p_{1/2}$&$  6s_{1/2}$ &  1780 &3.17[8]  &1.47[-1]  &1.70[ 0]\\
$5p_{1/2}$&$  7s_{1/2}$ &  1170 &8.18[7]  &1.65[-2]  &1.26[-1]\\
$5p_{1/2}$&$  8s_{1/2}$ &  1024 &7.29[6]  &1.13[-3]  &7.55[-3]\\
$5p_{3/2}$&$  6s_{1/2}$ &  1924 &5.76[8]  &1.56[-1]  &3.90[ 0]\\
$5p_{3/2}$&$  7s_{1/2}$ &  1231 &1.35[8]  &1.50[-2]  &2.41[-1]\\
$5p_{3/2}$&$  8s_{1/2}$ &  1070 &1.73[7]  &1.46[-3]  &2.04[-2]\\
$6p_{1/2}$&$  7s_{1/2}$ &  6859 &3.87[7]  &2.65[-1]  &1.18[ 1]\\
$6p_{1/2}$&$  8s_{1/2}$ &  3735 &1.45[7]  &3.00[-2]  &7.34[-1]\\
$6p_{3/2}$&$  7s_{1/2}$ &  7300 &7.38[7]  &2.86[-1]  &2.71[ 1]\\
$6p_{3/2}$&$  8s_{1/2}$ &  3862 &2.66[7]  &2.94[-2]  &1.49[ 0]\\
$7p_{1/2}$&$  8s_{1/2}$ & 15654 &1.07[7]  &3.79[-1]  &3.84[ 1]\\
$7p_{3/2}$&$  8s_{1/2}$ & 16592 &2.04[7]  &4.06[-1]  &8.72[ 1]\\
$6s_{1/2}$&$  6p_{1/2}$ &  6813 &5.89[7]  &4.14[-1]  &1.86[ 1]\\
$6s_{1/2}$&$  6p_{3/2}$ &  6428 &6.91[7]  &8.63[-1]  &3.67[ 1]\\
$6s_{1/2}$&$  7p_{1/2}$ &  2852 &1.76[5]  &2.15[-4]  &4.05[-3]\\
$6s_{1/2}$&$  7p_{3/2}$ &  2823 &1.30[6]  &3.11[-3]  &5.78[-2]\\
$6s_{1/2}$&$  8p_{1/2}$ &  2255 &2.52[5]  &1.93[-4]  &2.86[-3]\\
$6s_{1/2}$&$  8p_{3/2}$ &  2245 &1.38[1]  &2.08[-8]  &3.08[-7]\\
$7s_{1/2}$&$  7p_{1/2}$ & 17218 &1.21[7]  &5.71[-1]  &6.69[ 1]\\
$7s_{1/2}$&$  7p_{3/2}$ & 16210 &1.41[7]  &1.18[ 0]  &1.30[ 2]\\
$7s_{1/2}$&$  8p_{1/2}$ &  6625 &4.04[5]  &2.72[-3]  &1.20[-1]\\
$7s_{1/2}$&$  8p_{3/2}$ &  6545 &9.67[5]  &1.27[-2]  &5.54[-1]\\
$8s_{1/2}$&$  8p_{1/2}$ & 34495 &3.62[6]  &6.95[-1]  &1.64[ 2]\\
$8s_{1/2}$&$  8p_{3/2}$ & 32425 &4.26[6]  &1.43[ 0]  &3.16[ 2]\\
\end{tabular}
\end{ruledtabular}
\end{table}

The second-order Coulomb correction to the transition matrix
element in the case of $V^{N-1}$ DF potential is given by
\cite{dip3}
\begin{equation}\label{eq15}
Z_{wv}^{(2)}\!=\!\sum_{na}\frac{z_{an}(g_{wnva}-g_{wnav})}{\varepsilon
_{a}+\varepsilon _{v}-\varepsilon _{n}-\varepsilon _{w}}+\sum_{na}
\frac{(g_{wavn}-g_{wanv})z_{na}}{\varepsilon _{a}+\varepsilon
_{w}-\varepsilon _{n}-\varepsilon _{v}}.
\end{equation}
The second-order Breit corrections are obtained from
Eq.~(\ref{eq15}) by changing  $g_{ijkl}$ to $b_{ijkl}$ \cite{li-en}.
The third-order Coulomb correction is obtained from
Eqs.~(\ref{eq11}) and ~(\ref{eq13}) as
\begin{eqnarray}
Z_{wv}^{(3)}\! &=&\!\langle \Psi_{w}^{(2)}| Z|
\Psi_{v}^{(0)}\rangle\!
+\!\langle \Psi_{w}^{(0)}|Z| \Psi_{v}^{(2)}\rangle\!
+\!\langle \Psi_{w}^{(1)}|Z| \Psi_{v}^{(1)}\rangle
\nonumber\\
\label{eq16}
 &-&\frac{Z_{wv}^{(1)}}{2}\left[ \langle
 \Psi_{v}^{(1)}|\Psi_{v}^{(1)}\rangle +\langle
 \Psi_{w}^{(1)}|\Psi_{w}^{(1)}\rangle \right],
\end{eqnarray}
where the last term arises from the normalization condition. In
Ref.~\cite{equation}, contributions to $Z_{wv}^{(3)}$ were
presented in a following form:
\begin{equation}\label{eq17}
Z_{wv}^{(3)}=Z^{\rm RPA}+Z^{\rm BO}+Z^{\rm SR}+Z^{\rm
NORM}\,.
\end{equation}
The first term here corresponds to the well known random phase
approximation (RPA). Though RPA corresponds to the summation of
certain MBPT terms to all orders, it is possible to include it
here using the procedure described in Ref.~\cite{equation}. Next
term $Z^{\rm BO}$ corresponds to the correction  which arise from
substituting DF orbitals with Brueckner ones. The last two terms
in Eq.~(\ref{eq17}) describe structural radiation, $Z^{\rm SR}$,
and normalization, $Z^{\rm NORM}$, corrections.

In the all-order SD calculation,  we substitute the all-order SD
wave function $\Psi _{v}^{\rm SD}$ into the matrix element
expression given by Eq.~(\ref{eq11}) \cite{blundell-li}:
\begin{equation}\label{eq18}
Z_{wv}^{\rm SD}=\frac{z_{wv}+Z^{(a)}+ \cdots +Z^{(t)}}{\sqrt{(1+N_{w})(1+N_{v})}%
}\,,
\end{equation}
where $z_{wv}$ is the DF matrix element (\ref{eq14}) and the terms
$Z^{(k)}$, $k=a \cdots t$ are linear  or quadratic function of the
excitation coefficients introduced in Eq.~(\ref{eq8}).
Normalization terms $N_{v,w}$ are quadratic functions of the
excitation coefficients. This expression completely incorporates
$Z^{(3)}$ and certain sets of MBPT terms are summed to all
orders~\cite{blundell-li}. The part of the fourth-order correction
that is not included in the SD matrix element \eqref{eq18} was
recently discussed by \citet{der-4}, but we do not include it
here.

In Tables~\ref{tab-osc1} and~\ref{tab-osc2}, we present
theoretical transition rates $A_r$, oscillator strengths $f$, and
line strengths $S$ for E1 transitions between low-lying states of
In~I and Sn~II, respectively. These results are obtained by
combining all-order E1 amplitudes \eqref{eq18} in the length gauge
and theoretical energies $E^\text{SD}_\text{tot}$ from
Table~\ref{tab-esd} using well-known expressions (see, for
example, Ref.~\cite{nist}).

\begin{table}[tbh]
\caption{\label{tab-osc-com1} Oscillator strengths $f$
 and wavelengths  $\lambda$ (\AA) in In~I. The SD data
 ($f^{\mathrm{SD}}$) are compared with semi-empirical calculations ($f^{\mathrm{SE}}$)
  from Ref.~\protect\cite{migd76} and  experimental data ($f^{\mathrm{expt}}$)
  from Ref.~\protect\cite{penkin65}.}
\begin{ruledtabular}
\begin{tabular}{lllllll}
\multicolumn{1}{c}{Lower}& \multicolumn{1}{c}{Upper}&
\multicolumn{1}{c}{$\lambda^{\mathrm{SD}}$}&
\multicolumn{1}{c}{$\lambda^{\mathrm{expt.}}$ }&
\multicolumn{1}{c}{$f^{\mathrm{SD}}$ }&
\multicolumn{1}{c}{$f^{\mathrm{SE}}$ }&
\multicolumn{1}{c}{$f^{\mathrm{expt.}}$ }\\\hline
$5p_{1/2}$&$6s_{1/2}$& 4153& 4102&  0.133  & 0.137   & 0.14   \\
$5p_{3/2}$&$6s_{1/2}$& 4576& 4511&  0.142  & 0.153   & 0.15   \\[0.4pc]
$5p_{1/2}$&$7s_{1/2}$& 2792& 2754&  0.016  & 0.0158  & 0.017  \\
$5p_{3/2}$&$7s_{1/2}$& 2977& 2933&  0.015  & 0.161   & 0.017  \\[0.4pc]
$5p_{1/2}$&$8s_{1/2}$& 2493& 2460&  0.0054 & 0.00541 & 0.006  \\
$5p_{3/2}$&$8s_{1/2}$& 2639& 2602&  0.0051 & 0.00539 & 0.006  \\[0.4pc]
$5p_{1/2}$&$9s_{1/2}$& 2370& 2340&  0.0025 & 0.00256 & 0.0029 \\
$5p_{3/2}$&$9s_{1/2}$& 2502& 2468&  0.0024 & 0.00254 & 0.0026 \\[0.4pc]
$5p_{1/2}$&$5d_{3/2}$& 3045& 3039&  0.361  & 0.51    & 0.36   \\
$5p_{3/2}$&$5d_{3/2}$& 3266& 3259&  0.040  & 0.056   & 0.06   \\
$5p_{3/2}$&$5d_{5/2}$& 3264& 3256&  0.353  & 0.49    & 0.37   \\[0.4pc]
$5p_{1/2}$&$6d_{3/2}$& 2578& 2560&  0.072  & 0.11    & 0.043  \\
$5p_{3/2}$&$6d_{3/2}$& 2734& 2713&  0.0077 & 0.011   & 0.006  \\
$5p_{3/2}$&$6d_{5/2}$& 2734& 2710&  0.068  & 0.10    & 0.052  \\[0.4pc]
$5p_{1/2}$&$7d_{3/2}$& 2410& 2388&  0.026  & 0.039   & 0.006  \\
$5p_{3/2}$&$7d_{3/2}$& 2547& 2523&  0.0027 & 0.0033  & 0.0014 \\
$5p_{3/2}$&$7d_{5/2}$& 2546& 2521&  0.024  & 0.035   & 0.009  \\[0.4pc]
$5p_{1/2}$&$8d_{3/2}$& 2329& 2306&  0.012  & 0.017   & 0.0003 \\
$5p_{3/2}$&$8d_{3/2}$& 2456& 2432&  0.0013 & 0.0016  &        \\
$5p_{3/2}$&$8d_{5/2}$& 2456& 2439&  0.011  & 0.016   & 0.0013 \\
$6p_{1/2}$&$7s_{1/2}$&22594&     & 0.266   &   0.274  \\
$6p_{3/2}$&$7s_{1/2}$&24184&     & 0.279   &   0.287  \\[0.4pc]
$6p_{1/2}$&$8s_{1/2}$&11463&     & 0.0222  &   0.233  \\
$6p_{3/2}$&$8s_{1/2}$&11858&     & 0.0207  &   0.218  \\[0.4pc]
$6p_{1/2}$&$9s_{1/2}$& 9264&     & 0.00729 &   0.00764\\
$6p_{3/2}$&$9s_{1/2}$& 9520&     & 0.00664 &   0.00702\\[0.4pc]
$6s_{1/2}$&$6p_{1/2}$&13669&     & 0.402   &   0.467  \\
$6s_{1/2}$&$6p_{3/2}$&13146&     & 0.813   &   0.944  \\[0.4pc]
$6s_{1/2}$&$7p_{1/2}$& 7002&     & 0.0103  &   0.0110 \\
$6s_{1/2}$&$7p_{3/2}$& 6949&     & 0.0284  &   0.0207 \\[0.4pc]
$6s_{1/2}$&$8p_{1/2}$& 5806&     & 0.00206 &   0.00223\\
$6s_{1/2}$&$8p_{3/2}$& 5787&     & 0.00642 &   0.00704\\[0.4pc]
\end{tabular}
\end{ruledtabular}
\end{table}

\begin{table}[tbh]
\caption{\label{tab-tran-com-50} Transition probabilities $A$ (in
10$^7$~s$^{-1}$) and wavelengths  $\lambda$ (\AA) in Sn~II. Our SD
results are compared with experimental data from
Ref.~\protect\cite{alonso-00}. }
\begin{ruledtabular}
\begin{tabular}{llllll}
\multicolumn{1}{c}{Lower}& \multicolumn{1}{c}{Upper}&
\multicolumn{1}{c}{$\lambda^{\mathrm{SD}}$}&
\multicolumn{1}{c}{$\lambda^{\mathrm{expt}}$ }&
\multicolumn{1}{c}{$A^{\mathrm{SD}}$ }&
\multicolumn{1}{c}{$A^{\mathrm{expt}}$ }\\\hline
  $6s_{1/2}$&$   6p_{1/2}$&    6813 &  6844 &      5.89   &     5.8$\pm$1.1   \\
  $6p_{1/2}$&$   7s_{1/2}$&    6859 &  6761 &      3.87   &     4.2$\pm$0.1  \\
  $6s_{1/2}$&$   6p_{3/2}$&    6428 &  6453 &      6.91   &     5.2$\pm$1.0  \\
  $6p_{3/2}$&$   8s_{1/2}$&    3862 &  3841 &      2.66   &     2.5$\pm$0.5 \\
  $6p_{1/2}$&$   8s_{1/2}$&    3735 &  3715 &      1.45   &     1.8$\pm$0.3 \\
\end{tabular}
\end{ruledtabular}
\end{table}

\begin{table}[tbh]
\caption{\label{tab-life} Lifetimes ${\tau}$ in ns
 for the $nl$  levels
 in indium.
 The SD data are compared with experimental
results  from ($a$)--Ref.~\protect\cite{life-73},
($b$)--Ref.~\protect\cite{life-pra-83},
($c$)--Ref.~\protect\cite{life-jpb-83}, and
($d$)--Ref.~\protect\cite{life-astr-83}.}
\begin{ruledtabular}
\begin{tabular}{lll|ll}
\multicolumn{1}{c}{Level}            &
\multicolumn{1}{c}{$\tau^{\rm SD}$}  &
\multicolumn{1}{c|}{$\tau^{\rm expt}$}&
\multicolumn{1}{c}{Level}            &
\multicolumn{1}{c}{$\tau^{\rm SD}$}  \\
\hline
   $6s_{1/2}$&   7.04& 7.5$\pm$0.7$^a$    \\
   $7s_{1/2}$&  21.5 & 19.5$\pm$1.5$^c$;
                       19.5$\pm$1.5$^d$;
                       27$\pm$6$^b$    & $6p_{1/2}$&  69.7\\
   $8s_{1/2}$&  47.7 & 53$\pm$5$^c$;
                       55$\pm$6$^b$    & $7p_{1/2}$& 219. \\
   $9s_{1/2}$&  89.4 & 118$\pm$10$^c$;
                       104$\pm$12$^b$  & $7p_{3/2}$& 192. \\
   $5d_{3/2}$&   6.45& 6.3$\pm$0.5$^a$ & $8p_{1/2}$& 473. \\
   $5d_{5/2}$&   6.78& 7.6$\pm$0.5$^a$ & $8p_{3/2}$& 414. \\
   $6d_{3/2}$&  19.2 & 21$\pm$3$^a$    & $4f_{5/2}$&  70.4\\
   $6d_{5/2}$&  20.1 & 22$\pm$3$^a$;
                       18.6$\pm$1.5$^c$;
                       18.6$\pm$1.5$^d$& $4f_{7/2}$&  70.4\\
   $7d_{3/2}$&  42.0 & 50$\pm$5$^a$;
                       200$\pm$4$^b$   & $5f_{5/2}$& 125. \\
   $7d_{5/2}$&  44.0 &  50$\pm$5$^a$;
                       154$\pm$10$^c$;
                       147$\pm$10$^b$  & $5f_{7/2}$& 125. \\
   $8d_{3/2}$&  75.7 & 317$\pm$22$^c$     \\
   $8d_{5/2}$&  77.2 & 300$\pm$60$^c$;
                      238$\pm$20$^b$      \\
   $6p_{3/2}$&  63.7 &55.0$\pm$4$^d$      \\
\end{tabular}
\end{ruledtabular}
\end{table}

Calculation of the transition amplitudes provides another test of
the quality of atomic-structure calculations and another measure
of the size of the correlation corrections. In
Tables~\ref{tab-osc-com1} and~\ref{tab-tran-com-50}, we compare
our results with available experimental data. For convenience, we
also present theoretical and experimental wavelengths for all
transitions.  There is good agreement with experimental results
for the strongest lines of In. For Sn~II, agreement is also good
with exception of the $6s$~--~$6p_{3/2}$ transition where
experimental value is much smaller  than the calculated one. Note
that  the theory and experiment are in good agreement for the
$6s$~--~$6p_{1/2}$ transition.

We also use E1 transition rates to calculate the lifetimes of
low-lying levels of In~I and Sn~II. We compare these lifetimes
$\tau^{(\rm SD)}$ with available experimental measurements in
Table~\ref{tab-life}. For $7d_j$-levels, the measurements from
Refs.~\cite{life-73,life-pra-83,life-jpb-83} gave rather different
lifetimes. Our calculations support the shorter times  obtained in
Ref.~\cite{life-73}.

\section{Hyperfine constants for indium}

Calculations of hyperfine constants follow the same pattern as
calculations of E1 amplitudes, described in the previous section.
The value of the nuclear magnetic moment for $^{115}$In used here
is taken from \cite{web}. Hyperfine constants for another odd
isotope, $^{113}$In, can be obtained using the scaling  factor
0.99785, which is indistinguishable from unity within the accuracy
of the present theory. In contrast with dipole amplitudes
considered above, the hyperfine structure is sensitive to the wave
function at short distances and to very different types of
correlation corrections.

Table~\ref{tab-hyp} shows that SD method significantly improves DF
values of the hyperfine constants of the lowermost levels. It is
rather unusual that correlation correction to the hyperfine
structure constant of $5p_{3/2}$ level is so small. For other
$p_{3/2}$-levels, correlation corrections are comparable to the
initial DF contribution. This situation is more typical for other
atoms with $ns^2np_{3/2}$ configuration, such as Tl
\cite{DFKP98,kozlov}.

\begin{table}[tbh]
\caption{\label{tab-hyp} Hyperfine constants, $A$ (in MHz) for
$^{115}$In ($I$=9/2, $\mu$=5.5408~\protect\cite{web}).
  Dirac-Fock (DF) and all-order (SD) calculations are compared to
  experimental values from Ref.~\protect\cite{hyp-5s} - ($a$),
  Ref.~\protect\cite{hyp-57a} - ($b$), and
 Ref.~\protect\cite{hyp-57b} - ($c$).
 }
\begin{ruledtabular}
\begin{tabular}{rrrr|rrr}
\multicolumn{1}{c}{Level}
&\multicolumn{1}{c}{DF}
&\multicolumn{1}{c}{SD}
&\multicolumn{1}{c|}{Exper.}
&\multicolumn{1}{c}{Level}
&\multicolumn{1}{c}{DF}
&\multicolumn{1}{c}{SD}
\\[0.4pc]
 \hline
 $6s_{1/2}$ & 983.0  &  1812  &1685$^a$ &$5d_{3/2}$ & 4.365  & -11.48 \\
 $7s_{1/2}$ & 335.6  &  544.5 &         &$6d_{3/2}$ & 2.305  & -11.20 \\
 $8s_{1/2}$ & 153.6  &  240.8 &         &$7d_{3/2}$ & 1.275  & -7.692 \\
 $9s_{1/2}$ & 83.10  &  128.1 &         &$8d_{3/2}$ & 0.805  & -5.385 \\[0.4pc]
 $5p_{1/2}$ & 1780   &  2306  &2282$^b$ &$5d_{5/2}$ & 1.862  &  47.83 \\
 $6p_{1/2}$ & 222.7  &  263.2 &         &$6d_{5/2}$ & 0.981  &  30.81 \\
 $7p_{1/2}$ & 85.15  &  95.61 &         &$7d_{5/2}$ & 0.543  & 18.95  \\
 $8p_{1/2}$ & 41.90  &  45.97 &         &$8d_{5/2}$ & 0.342  & 12.59  \\[0.4pc]
 $5p_{3/2}$ & 267.8  &  262.4 &242.2$^c$&$4f_{5/2}$ & 0.0611 & 0.1871 \\
 $6p_{3/2}$ & 35.69  &  77.82 &         &$5f_{5/2}$ & 0.0316 & 0.1055 \\
 $7p_{3/2}$ & 13.71  &  30.83 &         &$4f_{7/2}$ & 0.0339 & 0.2293 \\
 $8p_{3/2}$ & 6.767  &  15.42 &         &$5f_{7/2}$ & 0.0176 & 0.1658 \\[0.4pc]
\end{tabular}
\end{ruledtabular}
\end{table}

\section{Conclusion}

Summarizing results of the previous sections we can make several
conclusions. We have seen that all-order SD calculations, when
converge, provide an improvement to the third order MBPT
calculation. Convergence of the SD-equations is hampered by the
existence of low-lying excitations from the uppermost core shell
$5s$. The lowest such excitation corresponds to configuration
$5s5p^2$ that has positive parity and primarily affects SD-equations
for the valence $d$-wave. Because of that, we were not able to solve
these equations for Sn~II. To avoid this problem one has to exclude
$5s$-electrons from the core and consider In~I and Sn~II as
three-electron systems. However, our attempt to treat In~I and Sn~II
as three-electron systems within second-order RMBPT for the valence
model space, as suggested in Refs.~\cite{bor-en}, resulted in rather
poor agreement with experimental spectra. We conclude that valence
correlations for atoms in question can not be accurately accounted
within model space approach. It would be interesting to perform
CI+MBPT calculations \cite{DFK96b}, but it goes beyond the scope of
the present paper.

Another interesting observation concerns the addition of the missing
part of the third-order term $E^{(3)}_\mathrm{extra}$ to the SD
results. It was suggested in Ref.~\cite{safr-na} to add this term,
so that all third-order terms are accounted for. For heavy
alkali-metal atoms omission of this term leads to significant
discrepancies of the all-order values with experiment. One can see
from Table~\ref{tab-esd}, that for atoms considered here this term
does not improve agreement with experimental energies. We have also
found that first- and second-order Breit corrections tend to cancel
each other in agreement with \cite{KPT00a}; final Breit corrections
are small and can be neglected within present accuracy of the
theory.

\begin{acknowledgments}
The work of M.S.S. was supported in part by National Science
Foundation Grant  No.\ PHY-0457078. M.G.K. acknowledges support from
Russian Foundation for Basic Research, grant No. 05-02-16914, grant
from Petersburg State Scientific Center, and thanks University of
Delaware for hospitality.
\end{acknowledgments}


\begin{thebibliography}{58}
\expandafter\ifx\csname
natexlab\endcsname\relax\def\natexlab#1{#1}\fi
\expandafter\ifx\csname bibnamefont\endcsname\relax
  \def\bibnamefont#1{#1}\fi
\expandafter\ifx\csname bibfnamefont\endcsname\relax
  \def\bibfnamefont#1{#1}\fi
\expandafter\ifx\csname citenamefont\endcsname\relax
  \def\citenamefont#1{#1}\fi
\expandafter\ifx\csname url\endcsname\relax
  \def\url#1{\texttt{#1}}\fi
\expandafter\ifx\csname
urlprefix\endcsname\relax\def\urlprefix{URL }\fi
\providecommand{\bibinfo}[2]{#2}
\providecommand{\eprint}[2][]{\url{#2}}

\bibitem[{\citenamefont{Migdalek}(1976)}]{migd76}
\bibinfo{author}{\bibfnamefont{J.}~\bibnamefont{Migdalek}},
  \bibinfo{journal}{Can.\ J.\ Phys.} \textbf{\bibinfo{volume}{54}},
  \bibinfo{pages}{118} (\bibinfo{year}{1976}).

\bibitem[{\citenamefont{Migdalek and Baylis}(1979)}]{migd79}
\bibinfo{author}{\bibfnamefont{J.}~\bibnamefont{Migdalek}} \bibnamefont{and}
  \bibinfo{author}{\bibfnamefont{W.~E.} \bibnamefont{Baylis}},
  \bibinfo{journal}{J.\ Phys.\ B} \textbf{\bibinfo{volume}{12}},
  \bibinfo{pages}{2595} (\bibinfo{year}{1979}).

\bibitem[{\citenamefont{Gruzdev and Afanaseva}(1978)}]{gruzdev78}
\bibinfo{author}{\bibfnamefont{P.~F.} \bibnamefont{Gruzdev}} \bibnamefont{and}
  \bibinfo{author}{\bibfnamefont{N.~V.} \bibnamefont{Afanaseva}},
  \bibinfo{journal}{Opt.\ Spectrosk.} \textbf{\bibinfo{volume}{44}},
  \bibinfo{pages}{514} (\bibinfo{year}{1978}).

\bibitem[{\citenamefont{Marcinck and Migdalek}(1994)}]{migd94}
\bibinfo{author}{\bibfnamefont{R.}~\bibnamefont{Marcinck}} \bibnamefont{and}
  \bibinfo{author}{\bibfnamefont{J.}~\bibnamefont{Migdalek}},
  \bibinfo{journal}{J.\ Phys.\ B} \textbf{\bibinfo{volume}{27}},
  \bibinfo{pages}{5587} (\bibinfo{year}{1994}).

\bibitem[{\citenamefont{Alonso-Medina and Colon}(2000)}]{alonso-00}
\bibinfo{author}{\bibfnamefont{A.}~\bibnamefont{Alonso-Medina}}
  \bibnamefont{and} \bibinfo{author}{\bibfnamefont{C.}~\bibnamefont{Colon}},
  \bibinfo{journal}{Phys.\ Scr.} \textbf{\bibinfo{volume}{61}},
  \bibinfo{pages}{646} (\bibinfo{year}{2000}).

\bibitem[{\citenamefont{Alonso-Medina et~al.}(2005)\citenamefont{Alonso-Medina,
  Colon, and Rivero}}]{alonso-05}
\bibinfo{author}{\bibfnamefont{A.}~\bibnamefont{Alonso-Medina}},
  \bibinfo{author}{\bibfnamefont{C.}~\bibnamefont{Colon}}, \bibnamefont{and}
  \bibinfo{author}{\bibfnamefont{C.}~\bibnamefont{Rivero}},
  \bibinfo{journal}{Phys.\ Scr.} \textbf{\bibinfo{volume}{71}},
  \bibinfo{pages}{154} (\bibinfo{year}{2005}).

\bibitem[{\citenamefont{Dzuba and Flambaum}(2005)}]{dzuba-05}
\bibinfo{author}{\bibfnamefont{V.~A.} \bibnamefont{Dzuba}} \bibnamefont{and}
  \bibinfo{author}{\bibfnamefont{V.~V.} \bibnamefont{Flambaum}},
  \bibinfo{journal}{Phys.\ Rev.\ A} \textbf{\bibinfo{volume}{71}},
  \bibinfo{pages}{52509} (\bibinfo{year}{2005}).

\bibitem[{\citenamefont{Andesen and S{\o}rensen}(1972)}]{life-73}
\bibinfo{author}{\bibfnamefont{T.}~\bibnamefont{Andesen}} \bibnamefont{and}
  \bibinfo{author}{\bibfnamefont{G.}~\bibnamefont{S{\o}rensen}},
  \bibinfo{journal}{Phys.\ Rev.\ A} \textbf{\bibinfo{volume}{5}},
  \bibinfo{pages}{2447} (\bibinfo{year}{1972}).

\bibitem[{\citenamefont{J{\"{o}}nsson et~al.}(1983)\citenamefont{J{\"{o}}nsson,
  Lundberg, and Svanberg}}]{life-pra-83}
\bibinfo{author}{\bibfnamefont{G.}~\bibnamefont{J{\"{o}}nsson}},
  \bibinfo{author}{\bibfnamefont{H.}~\bibnamefont{Lundberg}}, \bibnamefont{and}
  \bibinfo{author}{\bibfnamefont{S.}~\bibnamefont{Svanberg}},
  \bibinfo{journal}{Phys.\ Rev.\ A} \textbf{\bibinfo{volume}{27}},
  \bibinfo{pages}{2930} (\bibinfo{year}{1983}).

\bibitem[{\citenamefont{{M. A. Zaki Ewiss} et~al.}(1983)\citenamefont{{M. A.
  Zaki Ewiss}, Snoek, and D{\"{o}}nszelmann}}]{life-astr-83}
\bibinfo{author}{\bibnamefont{{M. A. Zaki Ewiss}}},
  \bibinfo{author}{\bibfnamefont{C.}~\bibnamefont{Snoek}}, \bibnamefont{and}
  \bibinfo{author}{\bibfnamefont{A.}~\bibnamefont{D{\"{o}}nszelmann}},
  \bibinfo{journal}{Astron.\ Astrophys} \textbf{\bibinfo{volume}{121}},
  \bibinfo{pages}{327} (\bibinfo{year}{1983}).

\bibitem[{\citenamefont{{M. A. Zaki Ewiss} and Snoek}(1983)}]{life-jpb-83}
\bibinfo{author}{\bibnamefont{{M. A. Zaki Ewiss}}} \bibnamefont{and}
  \bibinfo{author}{\bibfnamefont{C.}~\bibnamefont{Snoek}},
  \bibinfo{journal}{J.\ Phys.\ B} \textbf{\bibinfo{volume}{16}},
  \bibinfo{pages}{L153} (\bibinfo{year}{1983}).

\bibitem[{\citenamefont{Schectman et~al.}(2000)\citenamefont{Schectman, Cheng,
  Curtis, Federman, Fritts, and Irving}}]{life-00}
\bibinfo{author}{\bibfnamefont{R.~M.} \bibnamefont{Schectman}},
  \bibinfo{author}{\bibfnamefont{S.}~\bibnamefont{Cheng}},
  \bibinfo{author}{\bibfnamefont{L.~J.} \bibnamefont{Curtis}},
  \bibinfo{author}{\bibfnamefont{S.~R.} \bibnamefont{Federman}},
  \bibinfo{author}{\bibfnamefont{M.~C.} \bibnamefont{Fritts}},
  \bibnamefont{and} \bibinfo{author}{\bibfnamefont{R.~E.}
  \bibnamefont{Irving}}, \bibinfo{journal}{Astr.\ J.}
  \textbf{\bibinfo{volume}{542}}, \bibinfo{pages}{400} (\bibinfo{year}{2000}).

\bibitem[{\citenamefont{Eck et~al.}(1957)\citenamefont{Eck, Lurio, and
  Kusch}}]{hyp-57a}
\bibinfo{author}{\bibfnamefont{T.~G.} \bibnamefont{Eck}},
  \bibinfo{author}{\bibfnamefont{A.}~\bibnamefont{Lurio}}, \bibnamefont{and}
  \bibinfo{author}{\bibfnamefont{P.}~\bibnamefont{Kusch}},
  \bibinfo{journal}{Phys.\ Rev.} \textbf{\bibinfo{volume}{106}},
  \bibinfo{pages}{954} (\bibinfo{year}{1957}).

\bibitem[{\citenamefont{Eck and Kusch}(1957)}]{hyp-57b}
\bibinfo{author}{\bibfnamefont{T.~G.} \bibnamefont{Eck}} \bibnamefont{and}
  \bibinfo{author}{\bibfnamefont{P.}~\bibnamefont{Kusch}},
  \bibinfo{journal}{Phys.\ Rev.} \textbf{\bibinfo{volume}{106}},
  \bibinfo{pages}{958} (\bibinfo{year}{1957}).

\bibitem[{\citenamefont{Neijzen and D{\"{o}}nszelmann}(1980)}]{hyp-5s}
\bibinfo{author}{\bibfnamefont{J.~H.~M.} \bibnamefont{Neijzen}}
  \bibnamefont{and}
  \bibinfo{author}{\bibfnamefont{A.}~\bibnamefont{D{\"{o}}nszelmann}},
  \bibinfo{journal}{Physica C} \textbf{\bibinfo{volume}{98}},
  \bibinfo{pages}{235} (\bibinfo{year}{1980}).

\bibitem[{\citenamefont{Karlsson and {U. Litz\'{e}n}}(2001)}]{hyp-01}
\bibinfo{author}{\bibfnamefont{H.}~\bibnamefont{Karlsson}} \bibnamefont{and}
  \bibinfo{author}{\bibnamefont{{U. Litz\'{e}n}}}, \bibinfo{journal}{J.\ Phys.\
  B} \textbf{\bibinfo{volume}{34}}, \bibinfo{pages}{4475}
  (\bibinfo{year}{2001}).

\bibitem[{\citenamefont{Zaal et~al.}(1978)\citenamefont{Zaal, Hogervorst,
  Eliel, Bouma, and Blok}}]{laser-78}
\bibinfo{author}{\bibfnamefont{G.~J.} \bibnamefont{Zaal}},
  \bibinfo{author}{\bibfnamefont{W.}~\bibnamefont{Hogervorst}},
  \bibinfo{author}{\bibfnamefont{E.~R.} \bibnamefont{Eliel}},
  \bibinfo{author}{\bibfnamefont{J.}~\bibnamefont{Bouma}}, \bibnamefont{and}
  \bibinfo{author}{\bibfnamefont{J.}~\bibnamefont{Blok}}, \bibinfo{journal}{J.\
  Phys.\ B} \textbf{\bibinfo{volume}{16}}, \bibinfo{pages}{2821}
  (\bibinfo{year}{1978}).

\bibitem[{\citenamefont{Riyves et~al.}(2003)\citenamefont{Riyves, Kelman, {Yu.
  Zhmenyak}, and {Yu. Shpenik}}}]{laser-03}
\bibinfo{author}{\bibfnamefont{R.}~\bibnamefont{Riyves}},
  \bibinfo{author}{\bibfnamefont{V.}~\bibnamefont{Kelman}},
  \bibinfo{author}{\bibnamefont{{Yu. Zhmenyak}}}, \bibnamefont{and}
  \bibinfo{author}{\bibnamefont{{Yu. Shpenik}}}, \bibinfo{journal}{Rad.\ Phys.\
  Chem.} \textbf{\bibinfo{volume}{68}}, \bibinfo{pages}{269}
  (\bibinfo{year}{2003}).

\bibitem[{\citenamefont{Favilla et~al.}(2007)\citenamefont{Favilla, Barsanti,
  and Bicchi}}]{laser-07}
\bibinfo{author}{\bibfnamefont{E.}~\bibnamefont{Favilla}},
  \bibinfo{author}{\bibfnamefont{S.}~\bibnamefont{Barsanti}}, \bibnamefont{and}
  \bibinfo{author}{\bibfnamefont{P.}~\bibnamefont{Bicchi}},
  \bibinfo{journal}{Rad.\ Phys.\ Chem.} \textbf{\bibinfo{volume}{76}},
  \bibinfo{pages}{440} (\bibinfo{year}{2007}).

\bibitem[{\citenamefont{Safronova et~al.}(2006)\citenamefont{Safronova, Cowan,
  and Safronova}}]{ga-06}
\bibinfo{author}{\bibfnamefont{U.~I.} \bibnamefont{Safronova}},
  \bibinfo{author}{\bibfnamefont{T.~E.} \bibnamefont{Cowan}}, \bibnamefont{and}
  \bibinfo{author}{\bibfnamefont{M.~S.} \bibnamefont{Safronova}},
  \bibinfo{journal}{J.\ Phys.\ B} \textbf{\bibinfo{volume}{39}},
  \bibinfo{pages}{749} (\bibinfo{year}{2006}).

\bibitem[{\citenamefont{Safronova et~al.}(2005)\citenamefont{Safronova,
  Johnson, and Safronova}}]{tl-05}
\bibinfo{author}{\bibfnamefont{U.~I.} \bibnamefont{Safronova}},
  \bibinfo{author}{\bibfnamefont{W.~R.} \bibnamefont{Johnson}},
  \bibnamefont{and} \bibinfo{author}{\bibfnamefont{M.~S.}
  \bibnamefont{Safronova}}, \bibinfo{journal}{Phys.\ Rev.\ A}
  \textbf{\bibinfo{volume}{71}}, \bibinfo{pages}{52506} (\bibinfo{year}{2005}).

\bibitem[{\citenamefont{Dzuba et~al.}(1996{\natexlab{a}})\citenamefont{Dzuba,
  Flambaum, and Kozlov}}]{DFK96a}
\bibinfo{author}{\bibfnamefont{V.~A.} \bibnamefont{Dzuba}},
  \bibinfo{author}{\bibfnamefont{V.~V.} \bibnamefont{Flambaum}},
  \bibnamefont{and} \bibinfo{author}{\bibfnamefont{M.~G.}
  \bibnamefont{Kozlov}}, \bibinfo{journal}{JETP Lett.}
  \textbf{\bibinfo{volume}{63}}, \bibinfo{pages}{882}
  (\bibinfo{year}{1996}{\natexlab{a}}).

\bibitem[{\citenamefont{Dzuba et~al.}(1996{\natexlab{b}})\citenamefont{Dzuba,
  Flambaum, and Kozlov}}]{DFK96b}
\bibinfo{author}{\bibfnamefont{V.~A.} \bibnamefont{Dzuba}},
  \bibinfo{author}{\bibfnamefont{V.~V.} \bibnamefont{Flambaum}},
  \bibnamefont{and} \bibinfo{author}{\bibfnamefont{M.~G.}
  \bibnamefont{Kozlov}}, \bibinfo{journal}{Phys. Rev. A}
  \textbf{\bibinfo{volume}{54}}, \bibinfo{pages}{3948}
  (\bibinfo{year}{1996}{\natexlab{b}}).

\bibitem[{\citenamefont{Kozlov}(2004)}]{Koz04}
\bibinfo{author}{\bibfnamefont{M.~G.} \bibnamefont{Kozlov}},
  \bibinfo{journal}{Int. J. Quant. Chem.} \textbf{\bibinfo{volume}{100}},
  \bibinfo{pages}{336} (\bibinfo{year}{2004}).

\bibitem[{\citenamefont{Moore}(1971)}]{nist}
\bibinfo{author}{\bibfnamefont{C.~E.} \bibnamefont{Moore}},
  \emph{\bibinfo{title}{Atomic Energy Levels - v. III, NSRDS-NBS 35}}
  (\bibinfo{publisher}{U. S. Government Printing Office},
  \bibinfo{address}{Washington DC}, \bibinfo{year}{1971}).

\bibitem[{\citenamefont{Sucher}(1980)}]{sucher}
\bibinfo{author}{\bibfnamefont{J.}~\bibnamefont{Sucher}},
  \bibinfo{journal}{Phys.\ Rev.\ A} \textbf{\bibinfo{volume}{22}},
  \bibinfo{pages}{348} (\bibinfo{year}{1980}).

\bibitem[{\citenamefont{Goldstone}(1957)}]{goldstone}
\bibinfo{author}{\bibfnamefont{J.}~\bibnamefont{Goldstone}},
  \bibinfo{journal}{Proc.\ R.\ Soc.\ London Ser.\ A}
  \textbf{\bibinfo{volume}{239}}, \bibinfo{pages}{267} (\bibinfo{year}{1957}).

\bibitem[{\citenamefont{Johnson et~al.}(1987)\citenamefont{Johnson, Idress, and
  Sapirstein}}]{wrj-en}
\bibinfo{author}{\bibfnamefont{W.~R.} \bibnamefont{Johnson}},
  \bibinfo{author}{\bibfnamefont{M.}~\bibnamefont{Idress}}, \bibnamefont{and}
  \bibinfo{author}{\bibfnamefont{J.}~\bibnamefont{Sapirstein}},
  \bibinfo{journal}{Phys.\ Rev.\ A} \textbf{\bibinfo{volume}{35}},
  \bibinfo{pages}{3218} (\bibinfo{year}{1987}).

\bibitem[{\citenamefont{Johnson
  et~al.}(1988{\natexlab{a}})\citenamefont{Johnson, Blundell, and
  Sapirstein}}]{li-en}
\bibinfo{author}{\bibfnamefont{W.~R.} \bibnamefont{Johnson}},
  \bibinfo{author}{\bibfnamefont{S.~A.} \bibnamefont{Blundell}},
  \bibnamefont{and}
  \bibinfo{author}{\bibfnamefont{J.}~\bibnamefont{Sapirstein}},
  \bibinfo{journal}{Phys.\ Rev.\ A} \textbf{\bibinfo{volume}{37}},
  \bibinfo{pages}{2764} (\bibinfo{year}{1988}{\natexlab{a}}).

\bibitem[{\citenamefont{Blundell et~al.}(1990)\citenamefont{Blundell, Johnson,
  and Sapirstein}}]{tl-en}
\bibinfo{author}{\bibfnamefont{S.~A.} \bibnamefont{Blundell}},
  \bibinfo{author}{\bibfnamefont{W.~R.} \bibnamefont{Johnson}},
  \bibnamefont{and}
  \bibinfo{author}{\bibfnamefont{J.}~\bibnamefont{Sapirstein}},
  \bibinfo{journal}{Phys.\ Rev.\ A} \textbf{\bibinfo{volume}{42}},
  \bibinfo{pages}{3751} (\bibinfo{year}{1990}).

\bibitem[{\citenamefont{Blundell et~al.}(1989)\citenamefont{Blundell, Johnson,
  Liu, and Sapirstein}}]{blundell-li}
\bibinfo{author}{\bibfnamefont{S.~A.} \bibnamefont{Blundell}},
  \bibinfo{author}{\bibfnamefont{W.~R.} \bibnamefont{Johnson}},
  \bibinfo{author}{\bibfnamefont{Z.~W.} \bibnamefont{Liu}}, \bibnamefont{and}
  \bibinfo{author}{\bibfnamefont{J.}~\bibnamefont{Sapirstein}},
  \bibinfo{journal}{Phys.\ Rev.\ A} \textbf{\bibinfo{volume}{40}},
  \bibinfo{pages}{2233} (\bibinfo{year}{1989}).

\bibitem[{Liu()}]{Liu}
\bibinfo{note}{Z. W. Liu, Ph.D. thesis, Notre Dame University, 1989.}

\bibitem[{\citenamefont{Blundell et~al.}(1991)\citenamefont{Blundell, Johnson,
  and Sapirstein}}]{blundell-cs}
\bibinfo{author}{\bibfnamefont{S.~A.} \bibnamefont{Blundell}},
  \bibinfo{author}{\bibfnamefont{W.~R.} \bibnamefont{Johnson}},
  \bibnamefont{and}
  \bibinfo{author}{\bibfnamefont{J.}~\bibnamefont{Sapirstein}},
  \bibinfo{journal}{Phys.\ Rev.\ A} \textbf{\bibinfo{volume}{43}},
  \bibinfo{pages}{3407} (\bibinfo{year}{1991}).

\bibitem[{\citenamefont{Safronova
  et~al.}(1998{\natexlab{a}})\citenamefont{Safronova, Derevianko, and
  Johnson}}]{safr-na}
\bibinfo{author}{\bibfnamefont{M.~S.} \bibnamefont{Safronova}},
  \bibinfo{author}{\bibfnamefont{A.}~\bibnamefont{Derevianko}},
  \bibnamefont{and} \bibinfo{author}{\bibfnamefont{W.~R.}
  \bibnamefont{Johnson}}, \bibinfo{journal}{Phys.\ Rev.\ A}
  \textbf{\bibinfo{volume}{58}}, \bibinfo{pages}{1016}
  (\bibinfo{year}{1998}{\natexlab{a}}).

\bibitem[{\citenamefont{Safronova
  et~al.}(1999{\natexlab{a}})\citenamefont{Safronova, Johnson, and
  Derevianko}}]{safr-alk}
\bibinfo{author}{\bibfnamefont{M.~S.} \bibnamefont{Safronova}},
  \bibinfo{author}{\bibfnamefont{W.~R.} \bibnamefont{Johnson}},
  \bibnamefont{and}
  \bibinfo{author}{\bibfnamefont{A.}~\bibnamefont{Derevianko}},
  \bibinfo{journal}{Phys.\ Rev.\ A} \textbf{\bibinfo{volume}{60}},
  \bibinfo{pages}{4476} (\bibinfo{year}{1999}{\natexlab{a}}).

\bibitem[{\citenamefont{Safronova and Johnson}(2004)}]{gold}
\bibinfo{author}{\bibfnamefont{U.~I.} \bibnamefont{Safronova}}
  \bibnamefont{and} \bibinfo{author}{\bibfnamefont{W.~R.}
  \bibnamefont{Johnson}}, \bibinfo{journal}{Phys.\ Rev.\ A}
  \textbf{\bibinfo{volume}{69}}, \bibinfo{pages}{052511}
  (\bibinfo{year}{2004}).

\bibitem[{\citenamefont{Johnson
  et~al.}(1988{\natexlab{b}})\citenamefont{Johnson, Blundell, and
  Sapirstein}}]{Bspline}
\bibinfo{author}{\bibfnamefont{W.~R.} \bibnamefont{Johnson}},
  \bibinfo{author}{\bibfnamefont{S.~A.} \bibnamefont{Blundell}},
  \bibnamefont{and}
  \bibinfo{author}{\bibfnamefont{J.}~\bibnamefont{Sapirstein}},
  \bibinfo{journal}{Phys.\ Rev.\ A} \textbf{\bibinfo{volume}{37}},
  \bibinfo{pages}{307} (\bibinfo{year}{1988}{\natexlab{b}}).

\bibitem[{\citenamefont{Safronova
  et~al.}(1996{\natexlab{a}})\citenamefont{Safronova, Johnson, and
  Safronova}}]{be-en}
\bibinfo{author}{\bibfnamefont{M.~S.} \bibnamefont{Safronova}},
  \bibinfo{author}{\bibfnamefont{W.~R.} \bibnamefont{Johnson}},
  \bibnamefont{and} \bibinfo{author}{\bibfnamefont{U.~I.}
  \bibnamefont{Safronova}}, \bibinfo{journal}{Phys.\ Rev.\ A}
  \textbf{\bibinfo{volume}{53}}, \bibinfo{pages}{4036}
  (\bibinfo{year}{1996}{\natexlab{a}}).

\bibitem[{\citenamefont{Safronova et~al.}(1997)\citenamefont{Safronova,
  Johnson, and Safronova}}]{be3-en}
\bibinfo{author}{\bibfnamefont{M.~S.} \bibnamefont{Safronova}},
  \bibinfo{author}{\bibfnamefont{W.~R.} \bibnamefont{Johnson}},
  \bibnamefont{and} \bibinfo{author}{\bibfnamefont{U.~I.}
  \bibnamefont{Safronova}}, \bibinfo{journal}{J.\ Phys.\ B}
  \textbf{\bibinfo{volume}{30}}, \bibinfo{pages}{2375} (\bibinfo{year}{1997}).

\bibitem[{\citenamefont{Grant}(1970)}]{Gra70}
\bibinfo{author}{\bibfnamefont{I.~P.} \bibnamefont{Grant}},
  \bibinfo{journal}{Advances in Physics} \textbf{\bibinfo{volume}{19}},
  \bibinfo{pages}{747} (\bibinfo{year}{1970}).

\bibitem[{\citenamefont{Lindroth et~al.}(1989)\citenamefont{Lindroth,
  M{\aa}rtensson-Pendrill, Ynnerman, and \"{O}ster}}]{LMY89}
\bibinfo{author}{\bibfnamefont{E.}~\bibnamefont{Lindroth}},
  \bibinfo{author}{\bibfnamefont{A.-M.} \bibnamefont{M{\aa}rtensson-Pendrill}},
  \bibinfo{author}{\bibfnamefont{A.}~\bibnamefont{Ynnerman}}, \bibnamefont{and}
  \bibinfo{author}{\bibfnamefont{P.}~\bibnamefont{\"{O}ster}},
  \bibinfo{journal}{J. Phys. B} \textbf{\bibinfo{volume}{22}},
  \bibinfo{pages}{2447} (\bibinfo{year}{1989}).

\bibitem[{\citenamefont{Kozlov et~al.}()\citenamefont{Kozlov, Porsev, and
  Tupitsyn}}]{KPT00a}
\bibinfo{author}{\bibfnamefont{M.~G.} \bibnamefont{Kozlov}},
  \bibinfo{author}{\bibfnamefont{S.~G.} \bibnamefont{Porsev}},
  \bibnamefont{and} \bibinfo{author}{\bibfnamefont{I.~I.}
  \bibnamefont{Tupitsyn}}, \bibinfo{note}{arXiv:\eprint{physics/0004076}
  (2000)}.

\bibitem[{\citenamefont{Mohr}(1974{\natexlab{a}})}]{mohr1}
\bibinfo{author}{\bibfnamefont{P.~J.} \bibnamefont{Mohr}},
  \bibinfo{journal}{Ann.\ Phys.\ (N.Y.)} \textbf{\bibinfo{volume}{88}},
  \bibinfo{pages}{26} (\bibinfo{year}{1974}{\natexlab{a}}).

\bibitem[{\citenamefont{Mohr}(1974{\natexlab{b}})}]{mohr2}
\bibinfo{author}{\bibfnamefont{P.~J.} \bibnamefont{Mohr}},
  \bibinfo{journal}{Ann.\ Phys.\ (N.Y.)} \textbf{\bibinfo{volume}{88}},
  \bibinfo{pages}{52} (\bibinfo{year}{1974}{\natexlab{b}}).

\bibitem[{\citenamefont{Mohr}(1975)}]{mohr3}
\bibinfo{author}{\bibfnamefont{P.~J.} \bibnamefont{Mohr}},
  \bibinfo{journal}{Phys.\ Rev.\ Lett.} \textbf{\bibinfo{volume}{34}},
  \bibinfo{pages}{1050} (\bibinfo{year}{1975}).

\bibitem[{\citenamefont{Kozlov et~al.}(2001)\citenamefont{Kozlov, Porsev, and
  Johnson}}]{kozlov}
\bibinfo{author}{\bibfnamefont{M.~G.} \bibnamefont{Kozlov}},
  \bibinfo{author}{\bibfnamefont{S.~G.} \bibnamefont{Porsev}},
  \bibnamefont{and} \bibinfo{author}{\bibfnamefont{W.~R.}
  \bibnamefont{Johnson}}, \bibinfo{journal}{Phys.\ Rev.\ A}
  \textbf{\bibinfo{volume}{64}}, \bibinfo{pages}{052107}
  (\bibinfo{year}{2001}).

\bibitem[{\citenamefont{Safronova
  et~al.}(1996{\natexlab{b}})\citenamefont{Safronova, Johnson, and
  Safronova}}]{bor-en}
\bibinfo{author}{\bibfnamefont{M.~S.} \bibnamefont{Safronova}},
  \bibinfo{author}{\bibfnamefont{W.~R.} \bibnamefont{Johnson}},
  \bibnamefont{and} \bibinfo{author}{\bibfnamefont{U.~I.}
  \bibnamefont{Safronova}}, \bibinfo{journal}{Phys.\ Rev.\ A}
  \textbf{\bibinfo{volume}{54}}, \bibinfo{pages}{2850}
  (\bibinfo{year}{1996}{\natexlab{b}}).

\bibitem[{\citenamefont{Safronova
  et~al.}(1999{\natexlab{b}})\citenamefont{Safronova, Johnson, and
  Livingston}}]{bor-tr}
\bibinfo{author}{\bibfnamefont{U.~I.} \bibnamefont{Safronova}},
  \bibinfo{author}{\bibfnamefont{W.~R.} \bibnamefont{Johnson}},
  \bibnamefont{and} \bibinfo{author}{\bibfnamefont{A.~E.}
  \bibnamefont{Livingston}}, \bibinfo{journal}{Phys.\ Rev.\ A}
  \textbf{\bibinfo{volume}{60}}, \bibinfo{pages}{996}
  (\bibinfo{year}{1999}{\natexlab{b}}).

\bibitem[{\citenamefont{Safronova
  et~al.}(1998{\natexlab{b}})\citenamefont{Safronova, Johnson, and
  Safronova}}]{bor3}
\bibinfo{author}{\bibfnamefont{U.~I.} \bibnamefont{Safronova}},
  \bibinfo{author}{\bibfnamefont{W.~R.} \bibnamefont{Johnson}},
  \bibnamefont{and} \bibinfo{author}{\bibfnamefont{M.~S.}
  \bibnamefont{Safronova}}, \bibinfo{journal}{At.\ Data Nucl.\ Data Tables}
  \textbf{\bibinfo{volume}{69}}, \bibinfo{pages}{183}
  (\bibinfo{year}{1998}{\natexlab{b}}).

\bibitem[{\citenamefont{Safronova et~al.}(2002)\citenamefont{Safronova, Namba,
  Albritton, Johnson, and Safronova}}]{alum-en}
\bibinfo{author}{\bibfnamefont{U.~I.} \bibnamefont{Safronova}},
  \bibinfo{author}{\bibfnamefont{C.}~\bibnamefont{Namba}},
  \bibinfo{author}{\bibfnamefont{J.~R.} \bibnamefont{Albritton}},
  \bibinfo{author}{\bibfnamefont{W.~R.} \bibnamefont{Johnson}},
  \bibnamefont{and} \bibinfo{author}{\bibfnamefont{M.~S.}
  \bibnamefont{Safronova}}, \bibinfo{journal}{Phys.\ Rev.\ A}
  \textbf{\bibinfo{volume}{65}}, \bibinfo{pages}{022507}
  (\bibinfo{year}{2002}).

\bibitem[{\citenamefont{Safronova et~al.}(2003)\citenamefont{Safronova, Sataka,
  Albritton, Johnson, and Safronova}}]{alum-tr}
\bibinfo{author}{\bibfnamefont{U.~I.} \bibnamefont{Safronova}},
  \bibinfo{author}{\bibfnamefont{M.}~\bibnamefont{Sataka}},
  \bibinfo{author}{\bibfnamefont{J.~R.} \bibnamefont{Albritton}},
  \bibinfo{author}{\bibfnamefont{W.~R.} \bibnamefont{Johnson}},
  \bibnamefont{and} \bibinfo{author}{\bibfnamefont{M.~S.}
  \bibnamefont{Safronova}}, \bibinfo{journal}{At.\ Data Nucl.\ Data Tables}
  \textbf{\bibinfo{volume}{84}}, \bibinfo{pages}{1} (\bibinfo{year}{2003}).

\bibitem[{\citenamefont{Johnson et~al.}(1997)\citenamefont{Johnson, Safronova,
  and Safronova}}]{3elec}
\bibinfo{author}{\bibfnamefont{W.~R.} \bibnamefont{Johnson}},
  \bibinfo{author}{\bibfnamefont{M.~S.} \bibnamefont{Safronova}},
  \bibnamefont{and} \bibinfo{author}{\bibfnamefont{U.~I.}
  \bibnamefont{Safronova}}, \bibinfo{journal}{Phys.\ Scripta}
  \textbf{\bibinfo{volume}{56}}, \bibinfo{pages}{252} (\bibinfo{year}{1997}).

\bibitem[{\citenamefont{Blundell et~al.}(1987)\citenamefont{Blundell, Guo,
  Johnson, and Sapirstein}}]{equation}
\bibinfo{author}{\bibfnamefont{S.~A.} \bibnamefont{Blundell}},
  \bibinfo{author}{\bibfnamefont{D.~S.} \bibnamefont{Guo}},
  \bibinfo{author}{\bibfnamefont{W.~R.} \bibnamefont{Johnson}},
  \bibnamefont{and}
  \bibinfo{author}{\bibfnamefont{J.}~\bibnamefont{Sapirstein}},
  \bibinfo{journal}{At.\ Data and Nucl.\ Data Tables}
  \textbf{\bibinfo{volume}{37}}, \bibinfo{pages}{103} (\bibinfo{year}{1987}).

\bibitem[{\citenamefont{Johnson et~al.}(1996)\citenamefont{Johnson, Liu, and
  Sapirstein}}]{dip3}
\bibinfo{author}{\bibfnamefont{W.~R.} \bibnamefont{Johnson}},
  \bibinfo{author}{\bibfnamefont{Z.~W.} \bibnamefont{Liu}}, \bibnamefont{and}
  \bibinfo{author}{\bibfnamefont{J.}~\bibnamefont{Sapirstein}},
  \bibinfo{journal}{At.\ Data and Nucl.\ Data Tables}
  \textbf{\bibinfo{volume}{64}}, \bibinfo{pages}{279} (\bibinfo{year}{1996}).

\bibitem[{\citenamefont{Derevianko and Emmons}(2002)}]{der-4}
\bibinfo{author}{\bibfnamefont{A.}~\bibnamefont{Derevianko}} \bibnamefont{and}
  \bibinfo{author}{\bibfnamefont{E.~D.} \bibnamefont{Emmons}},
  \bibinfo{journal}{Phys.\ Rev.\ A} \textbf{\bibinfo{volume}{66}},
  \bibinfo{pages}{012503} (\bibinfo{year}{2002}).

\bibitem[{\citenamefont{Penkin and Shabanova}(1965)}]{penkin65}
\bibinfo{author}{\bibfnamefont{N.~P.} \bibnamefont{Penkin}} \bibnamefont{and}
  \bibinfo{author}{\bibfnamefont{L.~N.} \bibnamefont{Shabanova}},
  \bibinfo{journal}{Opt.\ Spectrosc.} \textbf{\bibinfo{volume}{18}},
  \bibinfo{pages}{504} (\bibinfo{year}{1965}).

\bibitem[{web()}]{web}
\bibinfo{note}{URL = http://www.webelements.com}.

\bibitem[{\citenamefont{Dzuba et~al.}(1998)\citenamefont{Dzuba, Flambaum,
  Kozlov, and Porsev}}]{DFKP98}
\bibinfo{author}{\bibfnamefont{V.~A.} \bibnamefont{Dzuba}},
  \bibinfo{author}{\bibfnamefont{V.~V.} \bibnamefont{Flambaum}},
  \bibinfo{author}{\bibfnamefont{M.~G.} \bibnamefont{Kozlov}},
  \bibnamefont{and} \bibinfo{author}{\bibfnamefont{S.~G.}
  \bibnamefont{Porsev}}, \bibinfo{journal}{Sov. Phys.--JETP}
  \textbf{\bibinfo{volume}{87}}, \bibinfo{pages}{885} (\bibinfo{year}{1998}).

\end{thebibliography}
\end{document}